\documentclass[reprint,floatfix,longbibliography,twocolumn,notitlepage,superscriptaddress,pra,aps]{revtex4-2}
\usepackage[utf8]{inputenc}

\usepackage{url}
\usepackage{placeins}
\usepackage{booktabs}
\usepackage{makecell}
\usepackage{array}
\usepackage[colorlinks=true, allcolors=blue]{hyperref}
\usepackage[title]{appendix}
\usepackage[protrusion=true, expansion=true]{microtype}

\usepackage{graphicx}
\graphicspath{ {./fig/} }
\usepackage{svg}
\usepackage{caption, subcaption}

\usepackage{amsmath, amssymb}
\usepackage{dsfont}
\usepackage{cancel}
\usepackage{bm}
\usepackage{siunitx}

\newtheorem{theorem}{Theorem}

\usepackage{physics2}
\usephysicsmodule{ab, braket}

\usepackage{algorithm}
\usepackage{algpseudocode}
    \makeatletter
    \renewcommand{\Function}[2]{%
      \csname ALG@cmd@\ALG@L @Function\endcsname{#1}{#2}%
      \def\jayden@currentfunction{#1}%
      \edef\@currentHref{function.\jayden@currentfunction}%
      \Hy@raisedlink{\hyper@anchorstart{\@currentHref}\hyper@anchorend}%
    }
    
    \newcommand{\funclabel}[1]{%
      \@bsphack
      \protected@write\@auxout{}{%
        \string\newlabel{#1}{%
          {{\textsc{\jayden@currentfunction}}}%
          {\thepage}%
          {}%
          {\@currentHref}{}%
        }%
      }%
      \@esphack
    }
    \makeatother

\renewcommand{\O}{\mathcal{O}}
\DeclareMathOperator{\expec}{{\mathbb{E}}}
\DeclareMathOperator{\trace}{{tr}}
\newcommand{\M}{{\mathcal{M}}}
\newcommand{\e}{{\varepsilon}}
\DeclareMathOperator{\var}{Var}
\DeclareMathOperator{\med}{median}
\newcommand{\eye}{\mathds{1}}

\usepackage{orcidlink}

\begin{document}

\title{Classical Shadows with Improved Median-of-Means Estimation}

\author{Winston Fu\,\orcidlink{0009-0000-1383-6454}}
\affiliation{Quantum Innovation Centre (Q.InC), Agency for Science, Technology and Research (A*STAR), 2 Fusionopolis Way, Innovis \#08-03, Singapore 138634, Republic of Singapore\looseness=-1}
\affiliation{Institute of High Performance Computing (IHPC), Agency for Science, Technology and Research (A*STAR), 1 Fusionopolis Way, \#16-16 Connexis, Singapore 138632, Republic of Singapore\looseness=-1}

\author{Dax Enshan Koh\,\orcidlink{0000-0002-8968-591X}}
\affiliation{Quantum Innovation Centre (Q.InC), Agency for Science, Technology and Research (A*STAR), 2 Fusionopolis Way, Innovis \#08-03, Singapore 138634, Republic of Singapore\looseness=-1}
\affiliation{Institute of High Performance Computing (IHPC), Agency for Science, Technology and Research (A*STAR), 1 Fusionopolis Way, \#16-16 Connexis, Singapore 138632, Republic of Singapore\looseness=-1}
\affiliation{Science, Mathematics and Technology Cluster, Singapore University of Technology and Design, 8 Somapah Road, Singapore 487372, Republic of Singapore\looseness=-1}

\author{Siong Thye Goh\,\orcidlink{0000-0001-7563-0961}}
\affiliation{Quantum Innovation Centre (Q.InC), Agency for Science, Technology and Research (A*STAR), 2 Fusionopolis Way, Innovis \#08-03, Singapore 138634, Republic of Singapore\looseness=-1}
\affiliation{Institute of High Performance Computing (IHPC), Agency for Science, Technology and Research (A*STAR), 1 Fusionopolis Way, \#16-16 Connexis, Singapore 138632, Republic of Singapore\looseness=-1}
\affiliation{Singapore Management University, 81 Victoria St, Singapore 188065, Republic of Singapore \looseness=-1}

\author{Jian Feng Kong\,\orcidlink{0000-0001-5980-4140}}
\affiliation{Quantum Innovation Centre (Q.InC), Agency for Science, Technology and Research (A*STAR), 2 Fusionopolis Way, Innovis \#08-03, Singapore 138634, Republic of Singapore\looseness=-1}
\affiliation{Institute of High Performance Computing (IHPC), Agency for Science, Technology and Research (A*STAR), 1 Fusionopolis Way, \#16-16 Connexis, Singapore 138632, Republic of Singapore\looseness=-1}

\begin{abstract}
The classical shadows protocol, introduced by Huang et al.~[Nat.~Phys.~16, 1050 (2020)], makes use of the median-of-means (MoM) estimator to efficiently estimate the expectation values of $M$ observables with failure probability $\delta$ using only $\mathcal{O}(\log(M/\delta))$ measurements. In their analysis, Huang et al.\ used loose constants in their asymptotic performance bounds for simplicity. However, the specific values of these constants can significantly affect the number of shots used in practical implementations. To address this, we studied a modified MoM estimator proposed by Minsker [PMLR 195, 5925 (2023)] that uses optimal constants and involves a U-statistic over the data set. For efficient estimation, we implemented two types of incomplete U-statistics estimators, the first based on random sampling and the second based on cyclically permuted sampling. We compared the performance of the original and modified estimators when used with the classical shadows protocol with single-qubit Clifford unitaries (Pauli measurements) for an Ising spin chain, and global Clifford unitaries (Clifford measurements) for the Greenberger–Horne–Zeilinger (GHZ) state. While the original estimator outperformed the modified estimators for Pauli measurements, the modified estimators showed improved performance over the original estimator for Clifford measurements. Our findings highlight the importance of tailoring estimators to specific measurement settings to optimize the performance of the classical shadows protocol in practical applications.
\end{abstract}

\maketitle
\section{Introduction}

Classical shadows is a protocol recently proposed by Huang, Kueng, and Preskill \cite{huangPredictingManyProperties2020} that builds on ideas from Aaronson's shadow tomography \cite{aaronsonShadowTomographyQuantum2018} and quantum state tomography \cite{gutaFastStateTomography2020, sugiyamaPrecisionGuaranteedQuantumTomography2013} to efficiently estimate properties of an unknown quantum state \(\rho\), without requiring a complete description of \(\rho\) to be learned. The protocol tackles the fundamental scaling problem of quantum systems, where learning a full description of the quantum system requires a number of measurements that scales exponentially with the number of qubits \cite{haah2017sample, o2016efficient}. By instead constructing an approximate classical description of the system, the expectation values of \(M\) observables can be predicted using only \(\O(\log M)\) measurements, independent of system size. Since its introduction, the protocol and its variants \cite{lukensBayesianAnalysisClassical2021,huangEfficientEstimationPauli2021a,
hadfield2021adaptive,
zhao2021fermionic,
hadfieldMeasurementsQuantumHamiltonians2022,
zhouHybridFrameworkEstimating2023,  akhtarScalableFlexibleClassical2023,
huClassicalShadowTomography2023, shivam2023classical, akhtar2023scalable,wan2023matchgate,
innocenti2023shadow,ippoliti2023operator,helsen2023thrifty,
grierPrincipalEigenstateClassical2024, sauvageClassicalShadowsSymmetries2024, buClassicalShadowsPauliinvariant2024, grierSampleoptimalClassicalShadows2024,west2024real,ippoliti2024classical,king2024triply,
de2024classical,becker2024classical,hearth2024efficient,cai2024biased,chen2024quantum,heyraud2024unified,akhtar2024dual,wang2024quantum,wu2024contractive,zhou2024efficient,cioli2024approximate,liu2024auxiliary,bertoni2024shallow,anselmetti2024classical,mao2024magic,caprotti2024optimizing} have found applications across diverse domains, including chemistry \cite{zhang2023composite,huang2024evaluating,avdic2024enhanced,wang2024shadow,avdic2024fewer}, machine learning \cite{huangProvablyEfficientMachine2022, lewisImprovedMachineLearning2024, jerbiShadowsQuantumMachine2024}, error mitigation \cite{zhaoGrouptheoreticErrorMitigation2024, seifShadowDistillationQuantum2023}, entanglement characterization \cite{elban2020mixed,neven2021symmetry,rath2021quantum}, and variational quantum algorithms  \cite{boyd2022training,sackAvoidingBarrenPlateaus2022,basheer2023alternating,nakaji2023measurement}, among others \cite{garcia2021quantum,helsen2021estimating,white2023filtering,ippoliti2024learnability,ruiz2024restoring,conrad2024chasing}. Moreover, noise-robust adaptations of the protocol have been introduced, allowing it to sustain efficient performance under realistic, noisy conditions \cite{chen2020robust,kohClassicalShadowsNoise2022,brieger2023stability, nguyen2023shadow,wuErrormitigatedFermionicClassical2024, rozon2024optimal,jnane2024quantum,farias2024robust,onorati2024noise}.

To estimate expectation values of observables using classical shadows, the median-of-means (MoM) estimator \cite{devroyeSubGaussianMeanEstimators2015, lugosiMeanEstimationRegression2019} is commonly employed, offering favorable scaling of sample complexity with respect to the failure probability of the protocol. In the original construction, Huang et al.\ chose conservative numerical constants for estimator bounds, which, although not affecting theoretical analyses, play a critical role in practical implementations \cite{struchalinExperimentalEstimationQuantum2021, zhangExperimentalQuantumState2021, levyClassicalShadowsQuantum2024, duttPracticalBenchmarkingRandomized2023}. Specifically, these constants directly affect the number of measurements required to achieve a desired level of accuracy in the estimation process. While theoretical analyses typically focus on large sample sizes where constants become negligible, they become critical in practical scenarios with limited measurements or resource constraints. Tighter constants in the MoM estimator allow for more precise estimates with fewer measurements, reducing overhead.

Recently, Minsker \cite{minskerEfficientMedianMeans2023, minskerUstatisticsGrowingOrder2023} presented optimal constants for the MoM estimator, and also proposed a modified version of the estimator with even tighter constants. In this study, we apply Minsker's results to classical shadows. Using ideas from incomplete U-statistics \cite{leeUStatisticsTheoryPractice2019}, we introduce practical implementations of Minsker's estimator which would otherwise be impossible to run for large datasets. Through numerical simulations, we benchmarked the performance of the original and modified estimators using an Ising spin chain and the Greenberger–Horne–Zeilinger (GHZ) state \cite{greenberger1989going} to measure their performance, using the single-qubit Clifford unitary ensemble and global Clifford unitary ensemble respectively. We found that while the original MoM estimator performed better in the former case, the modified version showed improvements in the latter case.

The rest of our manuscript is structured as follows. In \autoref{sec:classical_shadows}, we review the basics of the classical shadows protocol. In \autoref{sec:mom}, we present the original MoM estimator used in the protocol, followed by Minsker's modified estimator in \autoref{sec:mod_mom}. Next, in \autoref{sec:inc_mom}, we present two practical implementations of Minsker's estimator using incomplete U-statistics in the context of classical shadows. We then benchmark the estimators for the Ising spin chain in \autoref{sec:benchmark_ising} and GHZ state in \autoref{sec:benchmark_ghz}. Finally, in \autoref{sec:discussion}, we provide a discussion of the results.

\section{Results}
\subsection{Classical Shadows}
\label{sec:classical_shadows}

The purpose of the classical shadows protocol is to efficiently predict functions of a density matrix \(\rho\). Of particular importance are linear functions, such as expectation values \(\{o_i\}\) of a set of \(M\) observables \(\{O_i\}\), which can be expressed as:
\begin{equation}
    o_i(\rho) = \trace(O_i \rho) \quad 1 \leq i \leq M.
\end{equation}

The protocol works by applying a transformation \(\rho \mapsto U \rho U^\dagger\), where \(U\) is randomly selected from an ensemble of unitaries. We consider two commonly used ensembles: random global Clifford unitaries, and tensor products of random single-qubit Clifford unitaries. After the transformation, we perform a computational-basis measurement to obtain \(\ket*{\hat b} \in \{0, 1\}^r\), where \(r\) is the number of qubits. Using this, we define the shadow channel $\M$, defined as
\begin{equation}
    \M(\rho) = \expec \ab[U^\dagger \ketbra*{\hat b}{\hat b} U].
\end{equation}

We typically choose ensembles for which the object \(U^\dagger \ketbra*{\hat b}{\hat b} U\) can be efficiently stored classically. Assuming that the inverse map \(\M^{-1}\) exists, it can be shown that the classical snapshot \(\hat \rho = \M^{-1} (U^\dagger \ketbra*{\hat b}{\hat b} U)\) is an unbiased estimator of \(\rho\), i.e.,
\begin{equation}
    \rho = \expec[\hat \rho] = \expec[\M^{-1} (U^\dagger \ketbra*{\hat b}{\hat b} U)].
\end{equation}

While the inverse map \(\M^{-1}\) is not necessarily a valid quantum channel, it can still be applied during classical post-processing. Repeatedly sampling \(U\) from the ensemble of unitaries \(N\) times results in a classical shadow \(S\) of length \(N\):
\begin{equation}
    S(\rho; N) = \ab\{\hat \rho_1, \ldots, \hat \rho_N\}.
\end{equation}

\subsection{Median-of-Means Estimator}
\label{sec:mom}

Given \(M\) observables \(O_i\), which return \(o_i = \trace(O_i \rho)\) when acting on the system \(\rho\), we can construct an unbiased estimator \(\hat o_i\) of \(o_i\) using the set of classical shadows \(S(\rho; N)\):
\begin{equation}
    \hat o_i(N, 1) = \frac{1}{N}\sum_{j=1}^N\trace(O_i\hat \rho_j).
\end{equation}

From Chebyshev's inequality,
\begin{align}
    P(|\hat o_i(N, 1) - \trace(O_i \rho)|\geq \e) &\leq \frac{\var[\hat o_i]}{\e^2}\\
    &= \frac{\var[\trace(O_i \hat \rho_1)]}{N\e^2}\\
    &\equiv \frac{\delta}{M}.
    \label{eq:mean_bound}
\end{align}
This means that to achieve a failure probability below \(\delta\), the number of samples required scales as \(N \sim 1/\delta\). To circumvent this \(1/\delta\) dependence, the MoM estimator is often used:
\begin{align}
    \hat o_i(N, k) = \med\left\{ \hat o_i^{(1)}\left(\left\lfloor \frac{N}{k}\right\rfloor , 1\right), \ldots, \hat o_i^{(k)}\left(\left\lfloor\frac{N}{k}\right\rfloor, 1\right) \right\}.
\end{align}
This involves splitting the \(N\) snapshots into groups of size \(\lfloor N/k\rfloor\), and finding the mean of each group. Then, we take the median of these groups. This procedure is shown in \autoref{alg:mom}.

\begin{algorithm}[H]
    \caption{Median-of-means estimator for a list of data (\sc{Mom}).}
    \label{alg:mom}
    \begin{algorithmic}
        \State \textbf{Input:} \begin{itemize}
        \item $x_1, \ldots, x_N$: Array of data.
        \item $k$: The number of disjoint subsets.
        \end{itemize}
        \State \textbf{Output:} Estimated result.
        \Statex
    \end{algorithmic}
    
    \begin{algorithmic}[1]
        \Function{Mom}{$x_1, \ldots, x_N, k$}
            \funclabel{func:mom}
            \State $A \gets \{\}$
            \For{$i \gets 0, \lfloor\frac{N}{k}\rfloor-1$}
                \State $A \gets A \cup \{\frac{1}{k}\sum_{p=ik+1}^{k(i+1)} x_p\}$
            \EndFor
            \State \textbf{return} median($A$)
        \EndFunction
    \end{algorithmic}
\end{algorithm}

Let \(\hat \mu_\text{MoM}\) be an MoM estimator of \(N\) random variables \(X_i\), so
\begin{equation}
    \hat\mu_\text{MoM} = \med(\bar X_1, \ldots, \bar X_k),
\end{equation}
where \(\bar X_i\) denotes the mean of the group with size \(\lfloor N/k\rfloor\). The estimator \(\hat\mu_\text{MoM}\) obeys the inequality
\begin{equation}
    \label{eq:mom_gen_bound}
    P\ab(|\hat\mu_\text{MoM} - \mu| \geq C\sigma \sqrt{\frac{t}{N}}) \leq 2 e^{-t},
\end{equation}
where \(C\) can be taken to be  \(8e^2\) (see Appendix \ref{sec:mom_bound}). Using \(N = 34 \sigma^2 k/ \e^2,\ t = k/2\), Huang et al.\ \cite{huangPredictingManyProperties2020} obtain the bound
\begin{equation}
\label{eq:mom_huang_bound}
     P\ab(|\hat\mu_\text{MoM} - \mu| \geq \e) \leq 2 e^{-k/2}.
\end{equation}
However, Minsker \cite{minskerDistributedStatisticalEstimation2019, minskerUstatisticsGrowingOrder2023} showed that a tighter bound is possible.

\begin{theorem}[Theorem 2.1 of Minsker \cite{minskerUstatisticsGrowingOrder2023}.]
\label{thm:mom_tight_bound}
    \begin{equation}
    P\ab(|\hat\mu_{\mathrm{MoM}} - \mu| \geq \sigma \sqrt{\frac{t}{N}}) \leq 2 \exp\ab(-\frac t\pi (1 + o(1))),
    \end{equation}
where \(o(1)\) is a function that goes to \(0\) as \(k, N/k \to \infty\), uniformly over \(t \in [l_{k, N}, u_{k, N}]\) for any sequences \(l_{k, N} \gg k g^2(N/k)\) and \(u_{k, N} \ll k\), where $g$ satisfies the following inequality:
\begin{equation}
    g(m) \leq C\expec \ab|\frac{X_1 - \mu}{\sigma}|^q m^{-(q-2)/2}
\end{equation}
whenever \(\expec|X_1 - \mu|^q < \infty\) for some \(q \in (2, 3]\). 
\end{theorem}

This corresponds to \(C = \sqrt{\pi} + o(1)\) in \autoref{eq:mom_gen_bound}. In the context of quantum measurements, the condition \(\expec|X_1 - \mu|^q < \infty\) is trivially satisfied. Next, if \(\sqrt{k} g(N/k) \to 0\) as \(k, N \to \infty\), then the range of \(t\) becomes \(1 \leq t \ll k\). For classical shadows, we can choose \(k\) such that \(N \gg k\) and \(o(1) \ll 1\). Hence, we will ignore this factor in our subsequent analysis.

Given \(M\) quantities we want to estimate, we construct the estimators \(\hat\mu_{1}, \ldots, \hat\mu_{M}\). We want to find a value for \(t\) such that:
\begin{equation}
    \label{eq:t_value}
    P\ab(|\hat\mu_{i} - \mu_i| \geq \e\ \forall i) \leq \delta.
\end{equation}
Starting from \autoref{thm:mom_tight_bound}, we make the substitution \(t \mapsto \pi t\), \(\e = \sigma \sqrt{\pi t / N}\): 
\begin{equation}
    P\ab(|\hat\mu_{i} - \mu_i| \geq \e) \leq 2 e^{-t} = \frac{\delta}{M}.
\end{equation}
From the union bound,
\begin{align}
    P\ab( \bigcup_{i=1}^M |\hat\mu_{i} - \mu_i| \geq \e) &\leq \sum_{i=1}^M P\ab(|\hat\mu_{i} - \mu_i| \geq \e) \\
    &\leq \sum_{i=1}^M \frac{\delta}{M} \\ 
    &= \delta.
\end{align}

Therefore, \(t = \log(2M / \delta)\). To satisfy the condition \(t \ll k\), we will choose \(t = k / \log(k)\).

\subsection{Modified Median-of-Means for Classical Shadows}
\label{sec:mod_mom}

\begin{algorithm}[H]
    \caption{Median-of-means estimator with combinations (\sc{MomComb}).}
    \label{alg:momcomb}
    \begin{algorithmic}
        \State \textbf{Input:} \begin{itemize}
        \item $x_1, \ldots, x_N$: Array of data.
        \item $k$: Changes the number of initial disjoint subsets, where $kl$ is the number of subsets.
        \item $l$: Size of sets of sample averages.
        \end{itemize}
        \State \textbf{Output:} Estimated result.
        \Statex
    \end{algorithmic}
        
    \begin{algorithmic}[1]
        \Function{MomComb}{$x_1, \ldots, x_N, k, l$}
            \funclabel{func:momcomb}
            \State $A \gets \{\}$
            \For{$i \gets 0, \lfloor\frac{N}{kl}\rfloor-1$} \Comment{Split data into disjoint subsets, like in \Call{Mom}{}.}
                \State $A \gets A \cup \{\frac{1}{kl}\sum_{j=ikl+1}^{kl(i+1)} x_j\}$
            \EndFor
            
            \Statex
            \State $B \gets$ all possible subsets of size $l$ of $A$ \Comment{$B$ contains $\binom{kl}{l}$ sets of size $l$.}
            \State $C \gets \{\}$

            \ForAll{$J\in B$}
                \State $C \gets C \cup \{\frac1l \sum_{p=1}^{l}J_p\}$
            \EndFor
            \State \textbf{return} median($C$)
        \EndFunction
    \end{algorithmic}
\end{algorithm}

It was shown by Minsker \cite{minskerEfficientMedianMeans2023} that a modified, permutation-invariant version of \ref{func:mom} can give even tighter bounds. Similar to before, we first split our set of \(N\) data points \(\{X_1, \ldots, X_N\}\) into disjoint subsets of size \(n = kl\) and take the mean of each subset:
\begin{equation}
    G_1\cup \ldots \cup G_{n} \subseteq [N],\quad Z_j = \bar X_j = \frac{1}{|G_j|} \sum_{i\in G_j} X_i,
\end{equation}
where \([N] = \{1, \ldots, N\}\). For some set of integers \(J \subseteq [n]\) of cardinality \(|J| = l\), define \( \bar Z_J = \frac1l \sum_{j\in J} Z_j\). Then, to construct our estimator \(\hat \mu_{\text{comb}}\), we take the median of all possible \(\bar Z_J\):
\begin{align}
    \hat\mu_{\text{comb}} &\equiv \med(\bar Z_J, J\in \mathcal{A}_n^{(l)}), \\
    \mathcal{A}_n^{(l)} &= \{J \subset [n]: |J| = l\}.
\end{align}
This process is shown in \autoref{alg:momcomb}.

Making appropriate substitutions to Minsker's results, we get a bound corresponding to \(C = \sqrt{2} + o(1)\) in \autoref{eq:mom_gen_bound}---a tighter constant compared to \ref{func:mom}.
\begin{theorem}[Theorem 1 of Minsker \cite{minskerEfficientMedianMeans2023}.]
\label{thm:new_bound}
    Assume that \(\expec|X_1 - \mu|^{2 + a} < \infty\) for some \(a > 0\). Suppose that \(l = o(m^a)\) and let \(L(n, l)\) and \(M(n, l)\) be any sequences such that \(L(n, l) \gg \frac nl g^2(m)\) and \(M(n, l) \ll \frac{n}{l^2}\). Then for all \(L(n, l) \leq t \leq M(n, l)\),
    \begin{equation}
    P\ab(|\hat\mu_{\mathrm{MoM}} - \mu| \geq \sigma \sqrt{\frac{t}{N}}) \leq 3 \exp\ab(-\frac {t}{2(1+o(1))}),
    \end{equation}
    where \(m = \lfloor N/k \rfloor\), \(o(1) \to 0\) as \(l, k \to \infty\) uniformly over all \(t \in [L(n, l), M(n, l)]\).
\end{theorem}

A full treatment can be found in Minsker's \cite{minskerEfficientMedianMeans2023} work. Using the same analysis as \autoref{eq:t_value}, we choose \(t = \log (3M / \delta)\). Additionally, we will use the values \(l = \log(N/k),\ t = {n}/{l^2 \log(l)}\).

Importantly, the performance of the \ref{func:momcomb} estimator is bounded by that of the U-statistic:
\begin{multline}
    U_{n, l}(\rho'_-) = {n \choose l}^{-1} \sum_{J\in \mathcal{A}_n^{(l)}} \bigg(\rho'_- \big( \sqrt m (\bar Z_J - \mu - \sqrt{t/N}) \\
    - \expec \rho'_- \big) \geq - \sqrt{k} \expec \rho'_- \bigg),
\end{multline}
where \(\rho(x) = |x|\), \(\rho'_-\) is the left derivative of \(\rho\), and 
\begin{equation}
    \expec \rho'_- \equiv \expec \rho'_-\left( \sqrt m(\bar Z_J - \mu - \sqrt{t/N}) \right).
\end{equation}

One downside of this approach is its computational complexity of \(\O (k^l \log k)\), which can quickly become untenable. A practical implementation therefore requires sampling a subset \(D \subseteq \mathcal{A}_n^{(l)}\) of \(m\) elements. In the proceeding sections, we will
draw heavily on analysis by Lee \cite{leeUStatisticsTheoryPractice2019} to present and benchmark two possible methods of sampling: randomly, and using cyclic permutations.

\subsection{Incomplete Modified Median-of-Means}
\label{sec:inc_mom}

\begin{algorithm}[H]
    \caption{Median-of-means estimator with random sampling (\sc{MomRand}).}
    \label{alg:momrand}
    \begin{algorithmic}
        \State \textbf{Input:} \begin{itemize}
        \item $x_1, \ldots, x_N$: Array of data.
        \item $k$: Changes the number of initial disjoint subsets, where $kl$ is the number of subsets.
        \item $l$: Size of sets of sample averages.
        \item $m$: Number of randomly sampled sets of sample averages.
        \end{itemize}
        \State \textbf{Output:} Estimated result.
        \Statex
    \end{algorithmic}
    
    \begin{algorithmic}[1]
        \Function{MomRand}{$x_1, \ldots, x_N, k, l, m$}
            \funclabel{func:momrand}
             \State $A \gets \{\}$
            \For{$i \gets 0, \lfloor\frac{N}{kl}\rfloor-1$} \Comment{Split data into disjoint subsets, like in \Call{Mom}{}.}
                \State $A \gets A \cup \{\frac{1}{kl}\sum_{p=ikl+1}^{kl(i+1)} x_p\}$
            \EndFor

            \Statex
            \State $B \gets \{\}$
            \For{$i\gets 0, m-1$} \Comment{Randomly sample $m$ subsets of $A$.}
                \State $B\gets B \cup \{J\subseteq A : |J| = l\}$ where $J$ is a random subset of $A$.
            \EndFor

            \Statex
            \State $C \gets \{\}$
            \ForAll{$J\in B$}
                \State $C \gets C \cup \{\frac1l \sum_{p=1}^{l}J_p\}$
            \EndFor
            \State \textbf{return} median($C$)
        \EndFunction
    \end{algorithmic}
\end{algorithm}

An incomplete U-statistic \(U_n^{(0)}\) involves considering a subset \(\mathcal D\) of size \(m\) out of the \(n \choose l\) terms in \ref{func:momcomb}, where \(\mathcal D\) is commonly called the design of \(U_n^{(0)}\). For a given \(m\), we want to choose \(\mathcal D\) to minimise the variance of \(U_n^{(0)}\) to maximise the efficiency of the estimator.

A simple and common way of choosing the design is to randomly sample from \(\mathcal{A}_n^{(l)}\) either with or without replacement. For simplicity, we will consider the former. This approach is shown in \autoref{alg:momrand}. It can be shown that \cite{leeUStatisticsTheoryPractice2019}
\begin{equation}
    \var U_n^{(0)} = \frac{\sigma_l^2}{m} + (1 - m^{-1}) \var U_n,
\end{equation}
where for some U-statistic \(U_n = {n\choose l}^{-1} \sum_{(n, l)} \psi(X_{i_1}, \ldots, X_{i_l}) \) and the sum is taken over all permutations \( (i_1, \ldots, i_l) \) of \(\{ 1, 2, \ldots, l \}\),
\begin{align}
    \sigma_c^2 &= \var [ \psi_c(X_1, \ldots, X_c) ],
\end{align}
where
\begin{align}
    \psi_c(x_1, \ldots, x_c) &= \expec[\psi(X_1, \ldots, X_k) | X_1 = x_1, \ldots, X_c = x_c].
\end{align}

Let \(m = Kn\) for some constant \(K\). Comparing the variance with the complete estimator \(U_n\), we find the asymptotic relative efficiency (ARE)
\begin{align}
    \text{ARE} = \lim_{n \to \infty} \frac{\var U_n}{\var U_n^{(0)}} = \frac{Kl^2 \sigma_1^2}{l^2 \sigma_1^2 + \sigma_l^2}.
\end{align}

We can choose \(\mathcal D\) in a manner that is theoretically more efficient than simple random sampling. One possible minimum variance design employs cyclic permutations of our data set.
To our data set of \(n\) elements \(Z_j\), we apply the cyclic permutations
\begin{equation}
    \begin{pmatrix}
        1 & 2 & \ldots & n \\
        d_\nu \oplus 1 & d_\nu \oplus 2 & \ldots & d_\nu \oplus n
    \end{pmatrix}
\end{equation}
for \(\nu = 1, 2, \ldots, l\). The symbol \(\oplus\) denotes addition (mod \(n\)), and \(d_\nu\)'s are chosen such that \((d_\nu - d_{\nu'})\ (\text{mod } n)\) are distinct for any pair \((\nu, \nu') \) with \( \nu \neq \nu'\). In other words, the chosen \(d_\nu\) must form a modified Golomb ruler \cite{drakakisReviewAvailableConstruction2009}, which can be found in \(\O(l^2)\) time. Furthermore, the values of \(d_\nu\) can be precomputed and stored in a lookup table for reference during the execution of the estimator. Hence, this step has minimal impact on the performance of the algorithm.

After applying each cyclic permutation with offset \(d_\nu\), the elements are sequentially grouped into sets of size \(l\), with their average taken to give \(\bar Z_J\). Repeating the process \(K\) times with a new set of \(d_\nu\), where all pairs \(d_\nu\) used have distinct differences, gives a total of \(m = Kn\) values for \(\bar Z_J\). This is summarized in \autoref{alg:momcyc}.

This estimator \ref{func:momcyc} has variance
\begin{equation}
    \var U_n^{(0)} = m^{-1} \big( l (lK - 1)\sigma_1^2 + \sigma_l^2 \big)
\end{equation}
and
\begin{equation}
    \text{ARE} = \frac{Kl^2 \sigma_1^2}{l(lK - 1)\sigma_1^2 + \sigma_l^2},
\end{equation}
which is more efficient than \ref{func:momrand} for the same \(m\).

\begin{algorithm}[H]
    \caption{Median-of-means estimator with cyclic permutations (\sc{MomCyc}).}
    \label{alg:momcyc}
    \begin{algorithmic}
        \State \textbf{Input:} \begin{itemize}
        \item $x_1, \ldots, x_N$: Array of data.
        \item $k$: Changes the number of initial disjoint subsets, where $kl$ is the number of subsets.
        \item $l$: Size of sets of sample averages.
        \item $m$: Number of rounds.
        \end{itemize}
        \State \textbf{Output:} Estimated result.
        \State $\Rightarrow$ \Call{CycPerm}{$A$, $d$}: cyclically permutes all elements in $A$ by offset $d$.
        \State $\Rightarrow$ \Call{GolRul}{$l$, $D$}: generates a Golomb ruler $(\mathrm{mod}\ n)$ of length $l$, while also accounting for the differences stored in $D$.
        \Statex
    \end{algorithmic}
        
    \begin{algorithmic}[1]
        \Function{MomCyc}{$x_1, \ldots, x_N, k, l, m$}
            \funclabel{func:momcyc}
            \State $A \gets \{\}$
            \For{$i \gets 0, \lfloor\frac{N}{kl}\rfloor-1$} \Comment{Split data into disjoint subsets, like in \Call{Mom}{}.}
                \State $A \gets A \cup \{\frac{1}{kl}\sum_{p=ikl+1}^{kl(i+1)} x_p\}$ 
            \EndFor

            \Statex
            \State $B \gets \{\}$
            \State $D \gets \{\}$
            \For{$i \gets 0, m$}
                \State $C \gets$ \Call{GolRul}{$l, D$}
                \State $D \gets$ differences between any two elements in $C$.
                \ForAll{$d \in C$}
                    \State \Call{CycPerm}{$A, d$}
                    \For{$j \gets 0, \lfloor\frac{N}{kl^2}\rfloor-1$} \Comment{Split data into disjoint subsets of length $l$ and store the mean of each subset.}
                        \State $B \gets B \cup \{\frac{1}{l}\sum_{p=j+1}^{j+l}A_p\}$
                    \EndFor
                \EndFor
            \EndFor
            \State \textbf{return} median($B$)
        \EndFunction
    \end{algorithmic}
\end{algorithm}

\subsection{Benchmarking with Ising Chain}
\label{sec:benchmark_ising}

We first used classical shadows with random single-qubit Cliffords (hereby referred to as `Pauli measurements') to estimate the two-point correlator function \(\ab<\sigma_1^z \sigma_{i+1}^z>\) of a transverse field Ising model (TFIM), with Hamiltonian
\begin{equation}
    H = J \sum_i \sigma_i^Z \sigma_{i+1}^Z + h\sum_i \sigma_i^X.
\end{equation}

\begin{figure}
    \centering
    \includegraphics[width=\linewidth]{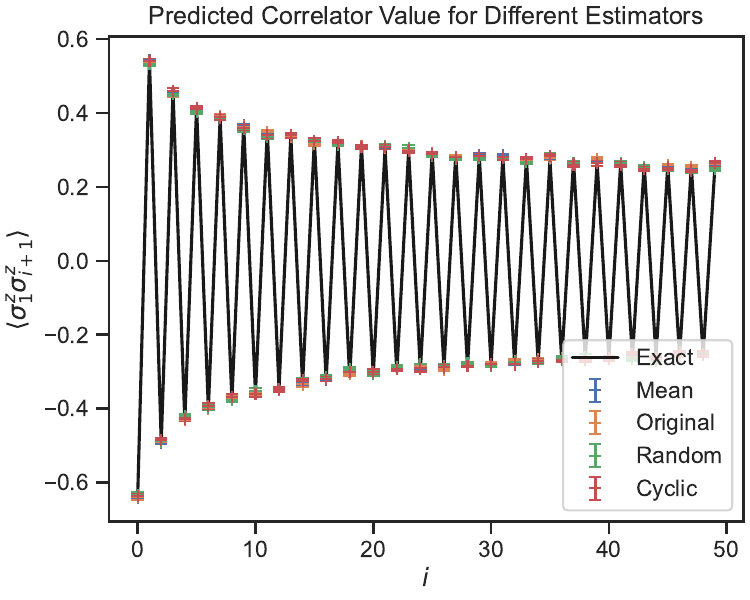}
    \caption{Predicted and exact values for the 2-point correlator function over 50 qubits, using Pauli measurements.}
    \label{fig:ising_pauli_results}
\end{figure}

\begin{figure}
    \centering
    \includegraphics[width=\linewidth]{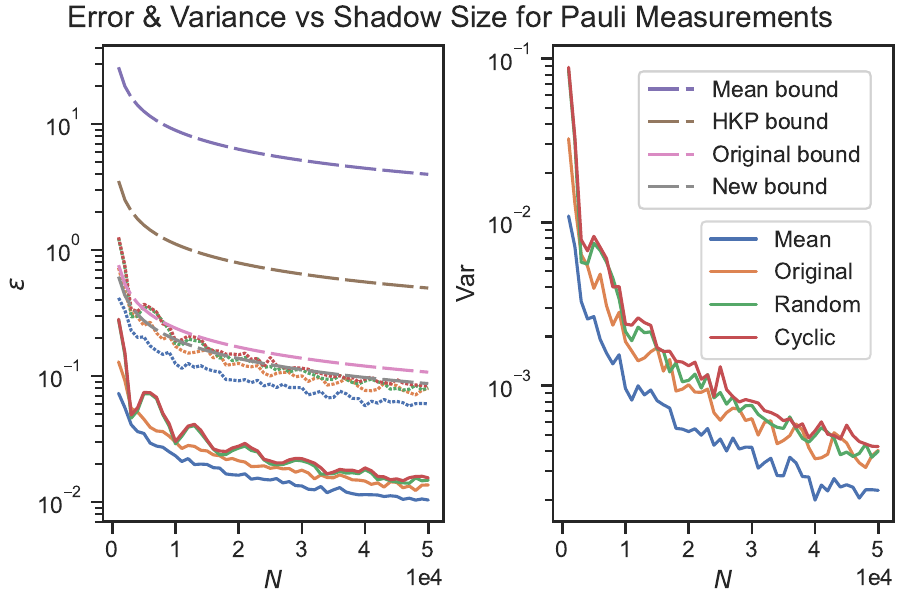}
    \caption{(Left) Average error of each correlator over 10 runs against shadow size for the mean, \ref{func:mom}, \ref{func:momcomb}, \ref{func:momrand}, and \ref{func:momcyc}. Short dotted lines of the corresponding color plot the \(3.3 \sigma\) boundary for each estimator. Long dashed lines show each of the bounds in \autoref{eq:mean_bound}, \autoref{thm:mom_tight_bound}, \autoref{thm:new_bound}, and \autoref{eq:mom_huang_bound}. (Right) Variance for the same runs.
    }
    \label{fig:ising_pauli_error}
\end{figure}

We found the correlator for 50 qubits with between \num{1000} and \num{50000} Pauli measurements, using tensor network techniques described in \autoref{sec:methods}. 

\autoref{fig:ising_pauli_results} shows the results of the simulations using 4 different kinds of estimators. \autoref{fig:ising_pauli_error}  shows the difference between the predicted and exact values, together with the `Mean bound' (\autoref{eq:mean_bound}), `Huang-Kueng-Preskill (HKP) bound' (\autoref{eq:mom_huang_bound}), `Original bound' (\autoref{thm:mom_tight_bound}), and `New bound' (\autoref{thm:new_bound}). 
The short dotted lines in the figure correspond to \(3.3 \sigma\) values, which have a \(0.1 \%\) probability of occurring. This allows for direct comparison between numerical results and the theoretical bounds (long dashed lines) with failure probability \(\delta = 0.1\%\).

\subsection{Benchmarking with Noisy GHZ State}
\label{sec:benchmark_ghz}

Next, we examined the fidelity between a noisy GHZ state and a pure GHZ state. We define
\begin{equation}
    \ket{\text{GHZ}^\pm_r} = \frac{1}{\sqrt{2}} (\ket 0 ^{\otimes r} \pm \ket 1 ^{\otimes r})
\end{equation}
and introduce a GHZ source which has a phase error with probability \(p\)
\begin{multline}
    \rho_p = (1-p) \ket{\text{GHZ}^+_r} \bra{\text{GHZ}^+_r} \\
    + p \ket{\text{GHZ}^-_r} \bra{\text{GHZ}^-_r}.
\end{multline}
Using Haar-random global Clifford unitaries (hereby referred to as `Clifford measurements'), we found the classical shadow of \(\rho_p\) and calculated the fidelity with the pure GHZ state \(\rho_{\text{pure}} = \ket{\text{GHZ}^+_r} \bra{\text{GHZ}^+_r}\),
\begin{equation}
    F(\rho_p, \rho_{\text{pure}}) = \trace(\rho_p\rho_{\text{pure}}).
\end{equation}

\begin{figure}
    \centering
    \includegraphics[width=\linewidth]{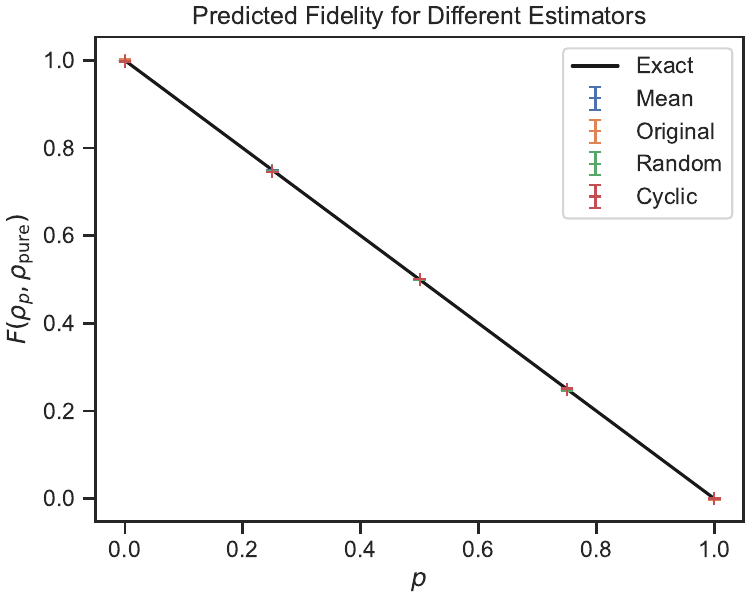}
    \caption{Predicted and exact values of fidelity for different phase error probability \(p\) using Clifford measurements.}
    \label{fig:ghz_results}
\end{figure}

\begin{figure}
    \centering
    \includegraphics[width=\linewidth]{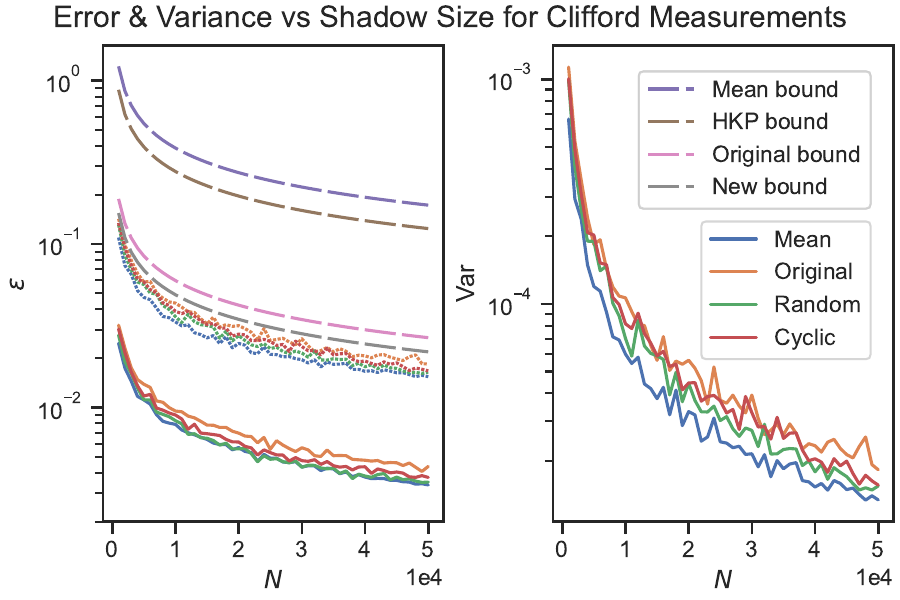}
    \caption{(Left) Average error for the fidelity over 100 runs against shadow size for the mean, \ref{func:mom}, \ref{func:momcomb}, \ref{func:momrand}, and \ref{func:momcyc} for the \(n=2\)-qubit noisy GHZ state. Short dotted lines of the corresponding color plot the \(3.3 \sigma\) boundary for each estimator. Long dashed lines show the bounds. (Right) Variance for the same runs.}
    \label{fig:ghz_error}
\end{figure}

\autoref{fig:ghz_results} shows the exact and predicted values of \(F(\rho_p, \rho_{\text{pure}})\) for difference estimators and \(r=2\). \autoref{fig:ghz_error} shows the average error and variance.

\begin{figure}
    \centering
    \includegraphics[width=1\linewidth]{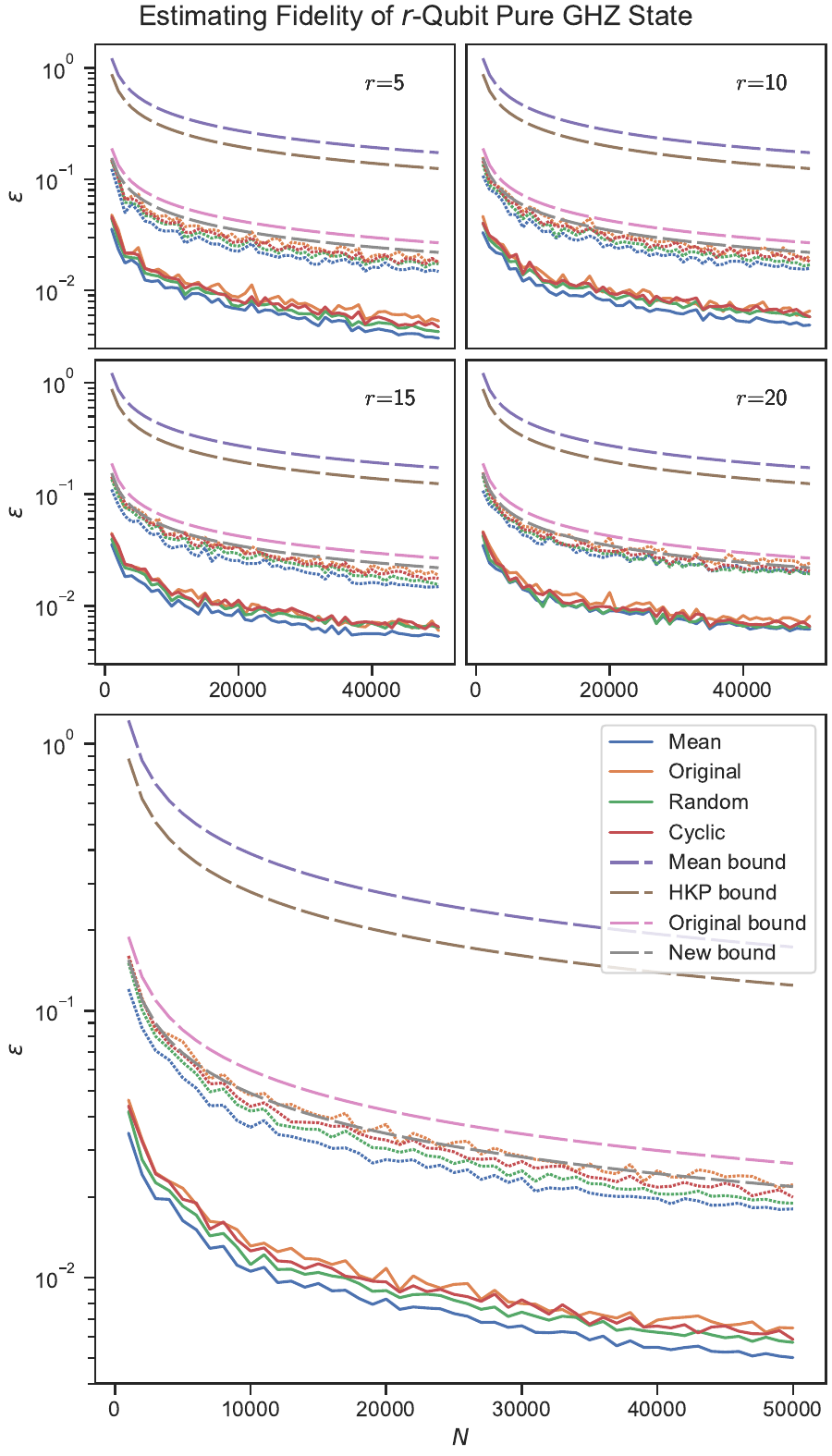}
    \caption{(Top half): Error in fidelity with the \(r\)-Qubit pure GHZ state with \(r=5, 10, 15, 20\), over 10 independent runs. (Bottom half): Average error over all four values of \(n\).}
    \label{fig:ghz_fid_var_qubits}
\end{figure}

\begin{figure}
    \centering
    \includegraphics[width=\linewidth]{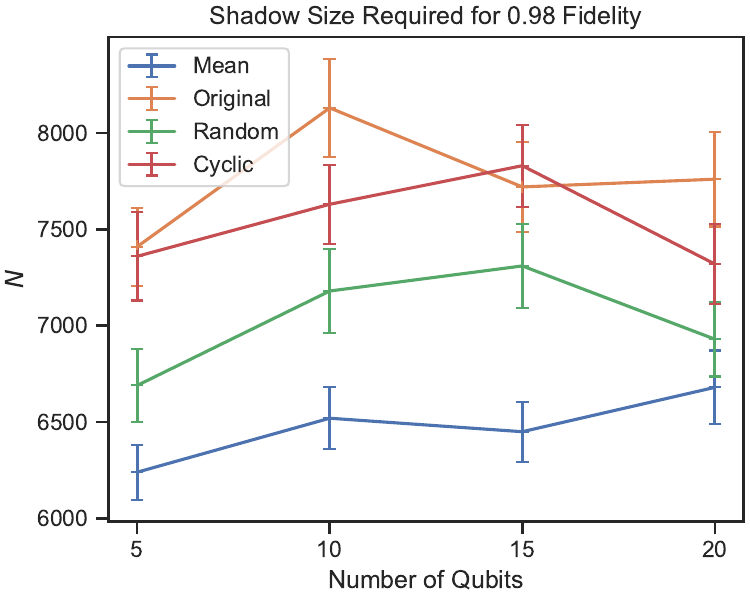}
    \caption{Number of snapshots needed to achieve 0.98 fidelity (\(\e < 0.02\)) for increasing system size, using various estimators, Averaged over 10 independent runs.}
    \label{fig:ghz_fid_var_qubits_2}
\end{figure}

Huang et al.\ \cite{huangPredictingManyProperties2020} showed that the performance of the classical shadows protocol is independent of the number of qubits in the system. To verify this for the new estimators, we estimated the fidelity of the pure \(p=0\) GHZ state for \(r=5, 10, 15, 20\), and the average. This is shown in \autoref{fig:ghz_fid_var_qubits}. In this case, the \(3.3\sigma\) value of \ref{func:mom} exceeds the bounds of the modified estimator (\autoref{thm:new_bound}). Next, the shadow size \(N\) required to attain a fidelity of \(0.98\) was found, as shown in \autoref{fig:ghz_fid_var_qubits_2}, showing independence between accuracy and shadow size regardless of estimator.

Finally, we tested the estimators on quadratic functions of \(\rho\). Instead of median of means, Huang et al.\ employ a median of U-statistics to estimate quadratic functions. The details of these modified estimators can be found in \autoref{sec:methods}.

\begin{figure}
    \centering
    \includegraphics[width=1\linewidth]{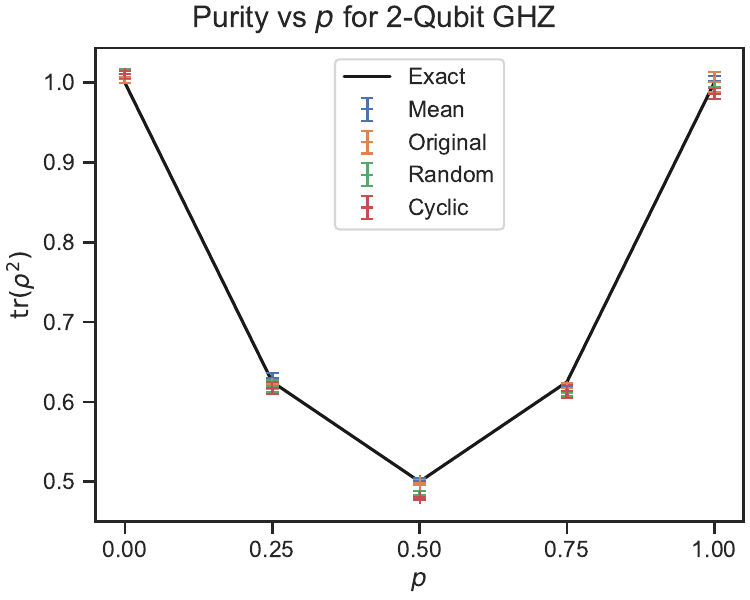}
    \caption{Predicted purity for the 2-qubits GHZ state using mean, \ref{func:mom}, \ref{func:momrand}, and \ref{func:momcyc}, for \(p=0, 0.25, 0.5, 0.75, 1\), and \(N=\num{10000}\) shadows.}
    \label{fig:ghz_purity}
\end{figure}

\begin{figure}
    \centering
    \includegraphics[width=1\linewidth]{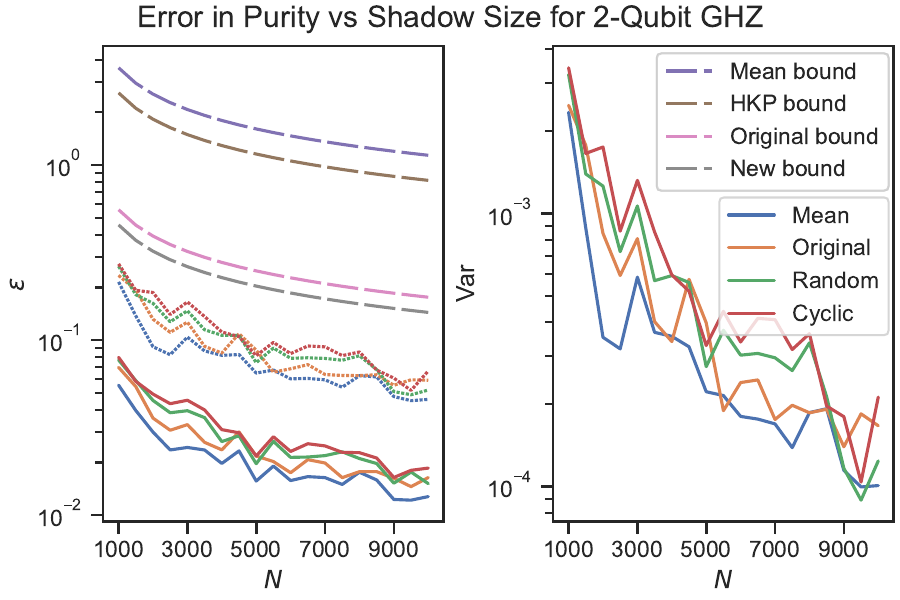}
    \caption{Error and variance of predictions with varying shadow size \(N\) using different estimators and their associated bounds, averaged over \(p=0, 0.25, 0.5, 0.75, 1\) and 10 independent runs.}
    \label{fig:ghz_purity_error}
\end{figure}

\autoref{fig:ghz_purity} shows the predicted purity for different values of \(p\) over 10 independent runs for \(p = 0, 0.25, 0.5, 0.75, 1\), and \autoref{fig:ghz_purity_error} shows the error and variances of each estimator.

\section{Discussion}
\label{sec:discussion}

Despite having the worst bound, taking the mean of the data had the lowest average error out of all the estimators. This agrees with prior experiments \cite{zhangExperimentalQuantumState2021, struchalinExperimentalEstimationQuantum2021, levyClassicalShadowsQuantum2024}, as the distribution of the estimator \(\hat o_i(N, 1)\) tends towards a normal distribution as \(N\to \infty\) and gives a tighter bound than \autoref{thm:new_bound}. Our results strengthen the case for the normality of \(\hat o_i(N, 1)\) for the number of samples \(N \geq 1000\) examined (Appendix \ref{sec:gaussian_bounds}).

When estimating Pauli observables using Pauli measurements, \ref{func:mom} had the best performance out of the remaining estimators despite having worse bounds than \ref{func:momrand} and \ref{func:momcyc}. \ref{func:momrand} had error that was in between \ref{func:mom} and \ref{func:momcyc}. This result is further discussed in Appendix \ref{sec:hit_count}. The \(3.3\sigma\) error of \ref{func:momrand} and \ref{func:momcyc} exceeded the tightened bounds by Minsker (\autoref{thm:new_bound}), indicating that these estimators may not be suitable for use with Pauli measurements.

When using Clifford measurements to predict linear functions, the practical performance of the MoM estimators followed the tightened bounds closely, indicating the validity of the assumptions made. \ref{func:mom} now had higher error and variance than \ref{func:momrand} and \ref{func:momcyc}. This shows that Minsker's estimator offers and advantage over the traditional MoM protocol for Clifford measurements. In this case, the mean and \ref{func:momrand} performed similarly, while \ref{func:momcyc} performed slightly worse. We have also demonstrated the tightness of Minsker's bound (\autoref{thm:new_bound}) in \autoref{fig:ghz_fid_var_qubits}, as the bound was exceeded by \ref{func:mom} but not \ref{func:momrand} and \ref{func:momcyc}.

When predicting quadratic functions such as purity, the mean continued to perform the best. The \ref{func:momrand} and \ref{func:momcyc} estimators had worse performance than \ref{func:mom}, in contrast to the improved performance for linear functions.

Despite the higher ARE of \ref{func:momcyc} compared to \ref{func:momrand}, the latter showed better performance for Clifford measurements.

While the mean showed the best performance in all the simulations, \ref{func:momrand} had similar results when using Clifford measurements, with the added benefit that the MoM estimators are backed by performance bounds under weaker assumptions. Using the bound in \autoref{thm:new_bound}, the number of shots needed can be greatly reduced. As a simple illustration, the minimum number of shots required is summarized in \autoref{tab:min_samples} for an average error \(M=50,\ \e < 0.1,\ \delta = 0.001\). The loose bound used by Huang et al.\ requires several orders of magnitude more shots than the tighter bounds by Minsker.

\begin{table}
    \centering
    \begin{tabular}{lr}
    \toprule
    Bound     & N \\
    \midrule
    Mean     & \num{80e6} \\
    HKP & \num{1.3e6} \\
    Original & \num{58e3} \\
    New & \num{38e3} \\
    \bottomrule
    \end{tabular}
    \caption{Number of samples required to achieve an average error of \(\e = 0.1\), using the different bounds.}
    \label{tab:min_samples}
\end{table}

As these changes happen purely in the classical post-processing step of the classical shadows protocol, our findings are immediately applicable to any experiment that employs the protocol with no changes to the actual experimental design, as well as on existing datasets.

\section{Methods}
\label{sec:methods}

\subsection{Pauli Measurements}
The ground state of the Ising Hamiltonian was found using the infinite time-evolving block decimation (iTEBD) algorithm, in four steps of bond dimensions \(\chi = 16, 32, 50, 100\) and time-steps \(\tau = 0.1, 0.01, 0.001, 0.0001\) respectively. This was treated as the ground truth, and the matrix product state (MPS) tensors were used to find the classical shadow. The bit-string \(\hat b = (\hat b_1, \ldots, \hat b_r) \in \{0, 1\}^r\) was found using methods proposed by Ferris and Vidal \cite{ferrisPerfectSamplingUnitary2012}. \num{50000} snapshots were taken for 10 independent experiments. Finally, the various estimators were used on the shadows with  \(M = 50,\ \delta = 0.001\). For \ref{func:momrand} and \ref{func:momcyc}, the number of samples was chosen to be \(m=10kl\).

To generate shadows of size between \num{1000} and \num{50000}, the appropriate number of snapshots was randomly sampled from the full shadow. The single-qubit Clifford unitary \(U_j\) applied to each qubit was stored and used to reconstruct each snapshot
\begin{equation}
    \label{eq:pauli_shadow_form}
    \hat \rho = \bigotimes_{j=1}^r\ab( 3 U_j^\dagger \ket*{\hat b_j} \bra*{\hat b_j} U_j - \eye ).
\end{equation}

\subsection{Clifford Measurements}

The GHZ state was simulated efficiently using the stabilizer formalism, implemented using the Qiskit library. Using the global Clifford ensemble, it can be shown that the snapshots have the form
\begin{equation}
    \hat \rho  = (2^r + 1)U^\dagger  \ket*{\hat b} \bra*{\hat b} U - \eye.
\end{equation}
When estimating fidelity we can use this form to show:
\begin{equation}
    F(\rho_0, \hat \rho) = (2^r + 1) |\bra*{\hat b} U \ket*{\text{GHZ}^+(r)}|^2 - 1.
\end{equation}
The shadows of the 2-qubit GHZ state were found for \(p=0, 0.25, 0.5, 0.75, 1\), for 100 independent runs and shadow size \(N = \num{50000}\). To estimate the fidelity for \(\num{1000} \leq N \leq \num{50000}\), the appropriate number of snapshots was sampled randomly from the full shadow.

Next, 10 shadows of the \(r=5, 10, 15, 20\) GHZ states were found for \(p=0\). For each shadow, 10 fidelity estimates were obtained for each value of \(\num{1000} \leq N \leq \num{50000}\), sampled randomly, using each of the estimators, to obtain 100 runs. To find the minimum number of snapshots for 0.98 fidelity, a rolling average over 5 values of \(N\) was used to smooth out the data, then the minimum \(N\) was found for each run, before taking the average.

To estimate purity,
\begin{equation}
    \trace(U_\text{swap} \hat\rho_1 \otimes \hat \rho_2) = (2^r+1)^2 |\bra*{\hat b_1} U_1 U_2^\dagger \ket*{\hat b_2} |^2 - 2 (2^r+1) + 2^r,
\end{equation}
where \(U_\text{swap}\) is the SWAP operator.
Instead of taking the median-of-means of disjoint subsets, we instead used each subset to find the U-statistic:
\begin{align}
    &\hat o_i^{(q)}\ab(\left\lfloor\frac Nk \right\rfloor, 1) \nonumber \\&= \frac{1}{\left\lfloor\frac Nk\right\rfloor(\left\lfloor\frac Nk\right\rfloor-1)} \sum_{\substack{u \neq v \\ u, v \in \{ N(1-\frac 1q)+1, \ldots, N\}}} \trace(O_i \hat\rho_u\otimes \hat\rho_v)
\end{align}
for \(1\leq q \leq k\). Then, we took the median of this set of U-statistics:
\begin{equation}
    \hat o_i(N, k) = \med \ab\{ \hat o_i^{(1)}\ab(\left\lfloor\frac Nk \right\rfloor, 1), \ldots, o_i^{(k)}\ab(\left\lfloor\frac Nk \right\rfloor, 1) \}.
\end{equation}
We define the `mean' estimator to be the mean of the set of \(\hat o_i^{(p)}(N, 1)\) from which the median is found for \ref{func:mom}.

To extend this estimator to \ref{func:momrand} and \ref{func:momcyc}, we treated this U-statistic as the mean in the first step of the algorithm (line 4 of Algorithms \ref{alg:momrand} and \ref{alg:momcyc}), then grouped them to form \(\bar Z_J\). This was used to estimate the purity \(\trace(\rho^2) = \trace(U_{\mathrm{swap}} \rho \otimes \rho)\) of the \(n=2\) noisy GHZ state for \(p=0, 0.25, 0.5, 0.75, 1\) from 10 shadows for \(1000 \leq N \leq 10000\).

\vspace{0.2cm}
\section*{Acknowledgements}
This research is supported by the National Research Foundation, Singapore, and the Agency for Science, Technology and Research (A*STAR), Singapore, under its Quantum Engineering Programme (NRF2021-QEP2-02-P03); A*STAR C230917003; and A*STAR under the Central Research Fund (CRF) Award for Use-Inspired Basic Research (UIBR) and the Quantum Innovation Centre (Q.InC) Strategic Research and Translational Thrust.

\bibliography{ref}

\begin{thebibliography}{89}%
\makeatletter
\providecommand \@ifxundefined [1]{%
 \@ifx{#1\undefined}
}%
\providecommand \@ifnum [1]{%
 \ifnum #1\expandafter \@firstoftwo
 \else \expandafter \@secondoftwo
 \fi
}%
\providecommand \@ifx [1]{%
 \ifx #1\expandafter \@firstoftwo
 \else \expandafter \@secondoftwo
 \fi
}%
\providecommand \natexlab [1]{#1}%
\providecommand \enquote  [1]{``#1''}%
\providecommand \bibnamefont  [1]{#1}%
\providecommand \bibfnamefont [1]{#1}%
\providecommand \citenamefont [1]{#1}%
\providecommand \href@noop [0]{\@secondoftwo}%
\providecommand \href [0]{\begingroup \@sanitize@url \@href}%
\providecommand \@href[1]{\@@startlink{#1}\@@href}%
\providecommand \@@href[1]{\endgroup#1\@@endlink}%
\providecommand \@sanitize@url [0]{\catcode `\\12\catcode `\$12\catcode `\&12\catcode `\#12\catcode `\^12\catcode `\_12\catcode `\%12\relax}%
\providecommand \@@startlink[1]{}%
\providecommand \@@endlink[0]{}%
\providecommand \url  [0]{\begingroup\@sanitize@url \@url }%
\providecommand \@url [1]{\endgroup\@href {#1}{\urlprefix }}%
\providecommand \urlprefix  [0]{URL }%
\providecommand \Eprint [0]{\href }%
\providecommand \doibase [0]{https://doi.org/}%
\providecommand \selectlanguage [0]{\@gobble}%
\providecommand \bibinfo  [0]{\@secondoftwo}%
\providecommand \bibfield  [0]{\@secondoftwo}%
\providecommand \translation [1]{[#1]}%
\providecommand \BibitemOpen [0]{}%
\providecommand \bibitemStop [0]{}%
\providecommand \bibitemNoStop [0]{.\EOS\space}%
\providecommand \EOS [0]{\spacefactor3000\relax}%
\providecommand \BibitemShut  [1]{\csname bibitem#1\endcsname}%
\let\auto@bib@innerbib\@empty
\bibitem [{\citenamefont {Huang}\ \emph {et~al.}(2020)\citenamefont {Huang}, \citenamefont {Kueng},\ and\ \citenamefont {Preskill}}]{huangPredictingManyProperties2020}%
  \BibitemOpen
  \bibfield  {author} {\bibinfo {author} {\bibfnamefont {H.-Y.}\ \bibnamefont {Huang}}, \bibinfo {author} {\bibfnamefont {R.}~\bibnamefont {Kueng}},\ and\ \bibinfo {author} {\bibfnamefont {J.}~\bibnamefont {Preskill}},\ }\bibfield  {title} {\bibinfo {title} {Predicting many properties of a quantum system from very few measurements},\ }\href {https://doi.org/10.1038/s41567-020-0932-7} {\bibfield  {journal} {\bibinfo  {journal} {Nature Physics}\ }\textbf {\bibinfo {volume} {16}},\ \bibinfo {pages} {1050} (\bibinfo {year} {2020})}\BibitemShut {NoStop}%
\bibitem [{\citenamefont {Aaronson}(2018)}]{aaronsonShadowTomographyQuantum2018}%
  \BibitemOpen
  \bibfield  {author} {\bibinfo {author} {\bibfnamefont {S.}~\bibnamefont {Aaronson}},\ }\bibfield  {title} {\bibinfo {title} {Shadow tomography of quantum states},\ }in\ \href {https://doi.org/10.1145/3188745.3188802} {\emph {\bibinfo {booktitle} {Proceedings of the 50th Annual ACM SIGACT Symposium on Theory of Computing}}},\ \bibinfo {series and number} {STOC 2018}\ (\bibinfo  {publisher} {Association for Computing Machinery},\ \bibinfo {address} {New York, NY, USA},\ \bibinfo {year} {2018})\ p.\ \bibinfo {pages} {325–338}\BibitemShut {NoStop}%
\bibitem [{\citenamefont {Gu{\c t}{\u a}}\ \emph {et~al.}(2020)\citenamefont {Gu{\c t}{\u a}}, \citenamefont {Kahn}, \citenamefont {Kueng},\ and\ \citenamefont {Tropp}}]{gutaFastStateTomography2020}%
  \BibitemOpen
  \bibfield  {author} {\bibinfo {author} {\bibfnamefont {M.}~\bibnamefont {Gu{\c t}{\u a}}}, \bibinfo {author} {\bibfnamefont {J.}~\bibnamefont {Kahn}}, \bibinfo {author} {\bibfnamefont {R.}~\bibnamefont {Kueng}},\ and\ \bibinfo {author} {\bibfnamefont {J.~A.}\ \bibnamefont {Tropp}},\ }\bibfield  {title} {\bibinfo {title} {Fast state tomography with optimal error bounds},\ }\href {https://doi.org/10.1088/1751-8121/ab8111} {\bibfield  {journal} {\bibinfo  {journal} {Journal of Physics A: Mathematical and Theoretical}\ }\textbf {\bibinfo {volume} {53}},\ \bibinfo {pages} {204001} (\bibinfo {year} {2020})}\BibitemShut {NoStop}%
\bibitem [{\citenamefont {Sugiyama}\ \emph {et~al.}(2013)\citenamefont {Sugiyama}, \citenamefont {Turner},\ and\ \citenamefont {Murao}}]{sugiyamaPrecisionGuaranteedQuantumTomography2013}%
  \BibitemOpen
  \bibfield  {author} {\bibinfo {author} {\bibfnamefont {T.}~\bibnamefont {Sugiyama}}, \bibinfo {author} {\bibfnamefont {P.~S.}\ \bibnamefont {Turner}},\ and\ \bibinfo {author} {\bibfnamefont {M.}~\bibnamefont {Murao}},\ }\bibfield  {title} {\bibinfo {title} {Precision-{{Guaranteed Quantum Tomography}}},\ }\href {https://doi.org/10.1103/PhysRevLett.111.160406} {\bibfield  {journal} {\bibinfo  {journal} {Physical Review Letters}\ }\textbf {\bibinfo {volume} {111}},\ \bibinfo {pages} {160406} (\bibinfo {year} {2013})}\BibitemShut {NoStop}%
\bibitem [{\citenamefont {Haah}\ \emph {et~al.}(2017)\citenamefont {Haah}, \citenamefont {Harrow}, \citenamefont {Ji}, \citenamefont {Wu},\ and\ \citenamefont {Yu}}]{haah2017sample}%
  \BibitemOpen
  \bibfield  {author} {\bibinfo {author} {\bibfnamefont {J.}~\bibnamefont {Haah}}, \bibinfo {author} {\bibfnamefont {A.}~\bibnamefont {Harrow}}, \bibinfo {author} {\bibfnamefont {Z.}~\bibnamefont {Ji}}, \bibinfo {author} {\bibfnamefont {X.}~\bibnamefont {Wu}},\ and\ \bibinfo {author} {\bibfnamefont {N.}~\bibnamefont {Yu}},\ }\bibfield  {title} {\bibinfo {title} {Sample-{O}ptimal {T}omography of {Q}uantum {S}tates},\ }\href {https://doi.org/10.1109/TIT.2017.2719044} {\bibfield  {journal} {\bibinfo  {journal} {IEEE Transactions on Information Theory}\ }\textbf {\bibinfo {volume} {63}},\ \bibinfo {pages} {5628} (\bibinfo {year} {2017})}\BibitemShut {NoStop}%
\bibitem [{\citenamefont {O'Donnell}\ and\ \citenamefont {Wright}(2016)}]{o2016efficient}%
  \BibitemOpen
  \bibfield  {author} {\bibinfo {author} {\bibfnamefont {R.}~\bibnamefont {O'Donnell}}\ and\ \bibinfo {author} {\bibfnamefont {J.}~\bibnamefont {Wright}},\ }\bibfield  {title} {\bibinfo {title} {Efficient quantum tomography},\ }in\ \href {https://doi.org/10.1145/2897518.2897544} {\emph {\bibinfo {booktitle} {Proceedings of the forty-eighth annual ACM symposium on Theory of Computing}}}\ (\bibinfo {year} {2016})\ pp.\ \bibinfo {pages} {899--912}\BibitemShut {NoStop}%
\bibitem [{\citenamefont {Lukens}\ \emph {et~al.}(2021)\citenamefont {Lukens}, \citenamefont {Law},\ and\ \citenamefont {Bennink}}]{lukensBayesianAnalysisClassical2021}%
  \BibitemOpen
  \bibfield  {author} {\bibinfo {author} {\bibfnamefont {J.~M.}\ \bibnamefont {Lukens}}, \bibinfo {author} {\bibfnamefont {K.~J.}\ \bibnamefont {Law}},\ and\ \bibinfo {author} {\bibfnamefont {R.~S.}\ \bibnamefont {Bennink}},\ }\bibfield  {title} {\bibinfo {title} {A {B}ayesian analysis of classical shadows},\ }\href {https://doi.org/10.1038/s41534-021-00447-6} {\bibfield  {journal} {\bibinfo  {journal} {npj Quantum Information}\ }\textbf {\bibinfo {volume} {7}},\ \bibinfo {pages} {113} (\bibinfo {year} {2021})}\BibitemShut {NoStop}%
\bibitem [{\citenamefont {Huang}\ \emph {et~al.}(2021{\natexlab{a}})\citenamefont {Huang}, \citenamefont {Kueng},\ and\ \citenamefont {Preskill}}]{huangEfficientEstimationPauli2021a}%
  \BibitemOpen
  \bibfield  {author} {\bibinfo {author} {\bibfnamefont {H.-Y.}\ \bibnamefont {Huang}}, \bibinfo {author} {\bibfnamefont {R.}~\bibnamefont {Kueng}},\ and\ \bibinfo {author} {\bibfnamefont {J.}~\bibnamefont {Preskill}},\ }\bibfield  {title} {\bibinfo {title} {Efficient {{Estimation}} of {{Pauli Observables}} by {{Derandomization}}},\ }\href {https://doi.org/10.1103/PhysRevLett.127.030503} {\bibfield  {journal} {\bibinfo  {journal} {Physical Review Letters}\ }\textbf {\bibinfo {volume} {127}},\ \bibinfo {pages} {030503} (\bibinfo {year} {2021}{\natexlab{a}})}\BibitemShut {NoStop}%
\bibitem [{\citenamefont {Hadfield}(2021)}]{hadfield2021adaptive}%
  \BibitemOpen
  \bibfield  {author} {\bibinfo {author} {\bibfnamefont {C.}~\bibnamefont {Hadfield}},\ }\bibfield  {title} {\bibinfo {title} {Adaptive {P}auli shadows for energy estimation},\ }\href {https://doi.org/10.48550/arXiv.2105.12207} {\bibfield  {journal} {\bibinfo  {journal} {arXiv preprint 2105.12207}\ } (\bibinfo {year} {2021})}\BibitemShut {NoStop}%
\bibitem [{\citenamefont {Zhao}\ \emph {et~al.}(2021)\citenamefont {Zhao}, \citenamefont {Rubin},\ and\ \citenamefont {Miyake}}]{zhao2021fermionic}%
  \BibitemOpen
  \bibfield  {author} {\bibinfo {author} {\bibfnamefont {A.}~\bibnamefont {Zhao}}, \bibinfo {author} {\bibfnamefont {N.~C.}\ \bibnamefont {Rubin}},\ and\ \bibinfo {author} {\bibfnamefont {A.}~\bibnamefont {Miyake}},\ }\bibfield  {title} {\bibinfo {title} {Fermionic partial tomography via classical shadows},\ }\href {https://doi.org/10.1103/PhysRevLett.127.110504} {\bibfield  {journal} {\bibinfo  {journal} {Physical Review Letters}\ }\textbf {\bibinfo {volume} {127}},\ \bibinfo {pages} {110504} (\bibinfo {year} {2021})}\BibitemShut {NoStop}%
\bibitem [{\citenamefont {Hadfield}\ \emph {et~al.}(2022)\citenamefont {Hadfield}, \citenamefont {Bravyi}, \citenamefont {Raymond},\ and\ \citenamefont {Mezzacapo}}]{hadfieldMeasurementsQuantumHamiltonians2022}%
  \BibitemOpen
  \bibfield  {author} {\bibinfo {author} {\bibfnamefont {C.}~\bibnamefont {Hadfield}}, \bibinfo {author} {\bibfnamefont {S.}~\bibnamefont {Bravyi}}, \bibinfo {author} {\bibfnamefont {R.}~\bibnamefont {Raymond}},\ and\ \bibinfo {author} {\bibfnamefont {A.}~\bibnamefont {Mezzacapo}},\ }\bibfield  {title} {\bibinfo {title} {Measurements of {{Quantum Hamiltonians}} with {{Locally-Biased Classical Shadows}}},\ }\href {https://doi.org/10.1007/s00220-022-04343-8} {\bibfield  {journal} {\bibinfo  {journal} {Communications in Mathematical Physics}\ }\textbf {\bibinfo {volume} {391}},\ \bibinfo {pages} {951} (\bibinfo {year} {2022})}\BibitemShut {NoStop}%
\bibitem [{\citenamefont {Zhou}\ and\ \citenamefont {Liu}(2024)}]{zhouHybridFrameworkEstimating2023}%
  \BibitemOpen
  \bibfield  {author} {\bibinfo {author} {\bibfnamefont {Y.}~\bibnamefont {Zhou}}\ and\ \bibinfo {author} {\bibfnamefont {Z.}~\bibnamefont {Liu}},\ }\bibfield  {title} {\bibinfo {title} {A hybrid framework for estimating nonlinear functions of quantum states},\ }\href {https://doi.org/10.1038/s41534-024-00846-5} {\bibfield  {journal} {\bibinfo  {journal} {npj Quantum Information}\ }\textbf {\bibinfo {volume} {10}},\ \bibinfo {pages} {62} (\bibinfo {year} {2024})}\BibitemShut {NoStop}%
\bibitem [{\citenamefont {Akhtar}\ \emph {et~al.}(2023{\natexlab{a}})\citenamefont {Akhtar}, \citenamefont {Hu},\ and\ \citenamefont {You}}]{akhtarScalableFlexibleClassical2023}%
  \BibitemOpen
  \bibfield  {author} {\bibinfo {author} {\bibfnamefont {A.~A.}\ \bibnamefont {Akhtar}}, \bibinfo {author} {\bibfnamefont {H.-Y.}\ \bibnamefont {Hu}},\ and\ \bibinfo {author} {\bibfnamefont {Y.-Z.}\ \bibnamefont {You}},\ }\bibfield  {title} {\bibinfo {title} {Scalable and {{Flexible Classical Shadow Tomography}} with {{Tensor Networks}}},\ }\href {https://doi.org/10.22331/q-2023-06-01-1026} {\bibfield  {journal} {\bibinfo  {journal} {Quantum}\ }\textbf {\bibinfo {volume} {7}},\ \bibinfo {pages} {1026} (\bibinfo {year} {2023}{\natexlab{a}})}\BibitemShut {NoStop}%
\bibitem [{\citenamefont {Hu}\ \emph {et~al.}(2023)\citenamefont {Hu}, \citenamefont {Choi},\ and\ \citenamefont {You}}]{huClassicalShadowTomography2023}%
  \BibitemOpen
  \bibfield  {author} {\bibinfo {author} {\bibfnamefont {H.-Y.}\ \bibnamefont {Hu}}, \bibinfo {author} {\bibfnamefont {S.}~\bibnamefont {Choi}},\ and\ \bibinfo {author} {\bibfnamefont {Y.-Z.}\ \bibnamefont {You}},\ }\bibfield  {title} {\bibinfo {title} {Classical shadow tomography with locally scrambled quantum dynamics},\ }\href {https://doi.org/10.1103/PhysRevResearch.5.023027} {\bibfield  {journal} {\bibinfo  {journal} {Physical Review Research}\ }\textbf {\bibinfo {volume} {5}},\ \bibinfo {pages} {023027} (\bibinfo {year} {2023})}\BibitemShut {NoStop}%
\bibitem [{\citenamefont {Shivam}\ \emph {et~al.}(2023)\citenamefont {Shivam}, \citenamefont {von Keyserlingk},\ and\ \citenamefont {Sondhi}}]{shivam2023classical}%
  \BibitemOpen
  \bibfield  {author} {\bibinfo {author} {\bibfnamefont {S.}~\bibnamefont {Shivam}}, \bibinfo {author} {\bibfnamefont {C.~W.}\ \bibnamefont {von Keyserlingk}},\ and\ \bibinfo {author} {\bibfnamefont {S.~L.}\ \bibnamefont {Sondhi}},\ }\bibfield  {title} {\bibinfo {title} {{On classical and hybrid shadows of quantum states}},\ }\href {https://doi.org/10.21468/SciPostPhys.14.5.094} {\bibfield  {journal} {\bibinfo  {journal} {SciPost Phys.}\ }\textbf {\bibinfo {volume} {14}},\ \bibinfo {pages} {094} (\bibinfo {year} {2023})}\BibitemShut {NoStop}%
\bibitem [{\citenamefont {Akhtar}\ \emph {et~al.}(2023{\natexlab{b}})\citenamefont {Akhtar}, \citenamefont {Hu},\ and\ \citenamefont {You}}]{akhtar2023scalable}%
  \BibitemOpen
  \bibfield  {author} {\bibinfo {author} {\bibfnamefont {A.~A.}\ \bibnamefont {Akhtar}}, \bibinfo {author} {\bibfnamefont {H.-Y.}\ \bibnamefont {Hu}},\ and\ \bibinfo {author} {\bibfnamefont {Y.-Z.}\ \bibnamefont {You}},\ }\bibfield  {title} {\bibinfo {title} {Scalable and {F}lexible {C}lassical {S}hadow {T}omography with {T}ensor {N}etworks},\ }\href {https://doi.org/10.22331/q-2023-06-01-1026} {\bibfield  {journal} {\bibinfo  {journal} {{Quantum}}\ }\textbf {\bibinfo {volume} {7}},\ \bibinfo {pages} {1026} (\bibinfo {year} {2023}{\natexlab{b}})}\BibitemShut {NoStop}%
\bibitem [{\citenamefont {Wan}\ \emph {et~al.}(2023)\citenamefont {Wan}, \citenamefont {Huggins}, \citenamefont {Lee},\ and\ \citenamefont {Babbush}}]{wan2023matchgate}%
  \BibitemOpen
  \bibfield  {author} {\bibinfo {author} {\bibfnamefont {K.}~\bibnamefont {Wan}}, \bibinfo {author} {\bibfnamefont {W.~J.}\ \bibnamefont {Huggins}}, \bibinfo {author} {\bibfnamefont {J.}~\bibnamefont {Lee}},\ and\ \bibinfo {author} {\bibfnamefont {R.}~\bibnamefont {Babbush}},\ }\bibfield  {title} {\bibinfo {title} {Matchgate shadows for fermionic quantum simulation},\ }\href {https://doi.org/10.1007/s00220-023-04844-0} {\bibfield  {journal} {\bibinfo  {journal} {Communications in Mathematical Physics}\ }\textbf {\bibinfo {volume} {404}},\ \bibinfo {pages} {629} (\bibinfo {year} {2023})}\BibitemShut {NoStop}%
\bibitem [{\citenamefont {Innocenti}\ \emph {et~al.}(2023)\citenamefont {Innocenti}, \citenamefont {Lorenzo}, \citenamefont {Palmisano}, \citenamefont {Albarelli}, \citenamefont {Ferraro}, \citenamefont {Paternostro},\ and\ \citenamefont {Palma}}]{innocenti2023shadow}%
  \BibitemOpen
  \bibfield  {author} {\bibinfo {author} {\bibfnamefont {L.}~\bibnamefont {Innocenti}}, \bibinfo {author} {\bibfnamefont {S.}~\bibnamefont {Lorenzo}}, \bibinfo {author} {\bibfnamefont {I.}~\bibnamefont {Palmisano}}, \bibinfo {author} {\bibfnamefont {F.}~\bibnamefont {Albarelli}}, \bibinfo {author} {\bibfnamefont {A.}~\bibnamefont {Ferraro}}, \bibinfo {author} {\bibfnamefont {M.}~\bibnamefont {Paternostro}},\ and\ \bibinfo {author} {\bibfnamefont {G.~M.}\ \bibnamefont {Palma}},\ }\bibfield  {title} {\bibinfo {title} {Shadow tomography on general measurement frames},\ }\href {https://doi.org/10.1103/PRXQuantum.4.040328} {\bibfield  {journal} {\bibinfo  {journal} {PRX Quantum}\ }\textbf {\bibinfo {volume} {4}},\ \bibinfo {pages} {040328} (\bibinfo {year} {2023})}\BibitemShut {NoStop}%
\bibitem [{\citenamefont {Ippoliti}\ \emph {et~al.}(2023)\citenamefont {Ippoliti}, \citenamefont {Li}, \citenamefont {Rakovszky},\ and\ \citenamefont {Khemani}}]{ippoliti2023operator}%
  \BibitemOpen
  \bibfield  {author} {\bibinfo {author} {\bibfnamefont {M.}~\bibnamefont {Ippoliti}}, \bibinfo {author} {\bibfnamefont {Y.}~\bibnamefont {Li}}, \bibinfo {author} {\bibfnamefont {T.}~\bibnamefont {Rakovszky}},\ and\ \bibinfo {author} {\bibfnamefont {V.}~\bibnamefont {Khemani}},\ }\bibfield  {title} {\bibinfo {title} {Operator relaxation and the optimal depth of classical shadows},\ }\href {https://doi.org/10.1103/PhysRevLett.130.230403} {\bibfield  {journal} {\bibinfo  {journal} {Phys. Rev. Lett.}\ }\textbf {\bibinfo {volume} {130}},\ \bibinfo {pages} {230403} (\bibinfo {year} {2023})}\BibitemShut {NoStop}%
\bibitem [{\citenamefont {Helsen}\ and\ \citenamefont {Walter}(2023)}]{helsen2023thrifty}%
  \BibitemOpen
  \bibfield  {author} {\bibinfo {author} {\bibfnamefont {J.}~\bibnamefont {Helsen}}\ and\ \bibinfo {author} {\bibfnamefont {M.}~\bibnamefont {Walter}},\ }\bibfield  {title} {\bibinfo {title} {Thrifty shadow estimation: Reusing quantum circuits and bounding tails},\ }\href {https://doi.org/10.1103/PhysRevLett.131.240602} {\bibfield  {journal} {\bibinfo  {journal} {Phys. Rev. Lett.}\ }\textbf {\bibinfo {volume} {131}},\ \bibinfo {pages} {240602} (\bibinfo {year} {2023})}\BibitemShut {NoStop}%
\bibitem [{\citenamefont {Grier}\ \emph {et~al.}(2024{\natexlab{a}})\citenamefont {Grier}, \citenamefont {Pashayan},\ and\ \citenamefont {Schaeffer}}]{grierPrincipalEigenstateClassical2024}%
  \BibitemOpen
  \bibfield  {author} {\bibinfo {author} {\bibfnamefont {D.}~\bibnamefont {Grier}}, \bibinfo {author} {\bibfnamefont {H.}~\bibnamefont {Pashayan}},\ and\ \bibinfo {author} {\bibfnamefont {L.}~\bibnamefont {Schaeffer}},\ }\bibfield  {title} {\bibinfo {title} {Principal eigenstate classical shadows},\ }in\ \href {https://proceedings.mlr.press/v247/grier24a.html} {\emph {\bibinfo {booktitle} {Proceedings of Thirty Seventh Conference on Learning Theory}}},\ \bibinfo {series} {Proceedings of Machine Learning Research}, Vol.\ \bibinfo {volume} {247},\ \bibinfo {editor} {edited by\ \bibinfo {editor} {\bibfnamefont {S.}~\bibnamefont {Agrawal}}\ and\ \bibinfo {editor} {\bibfnamefont {A.}~\bibnamefont {Roth}}}\ (\bibinfo  {publisher} {PMLR},\ \bibinfo {year} {2024})\ pp.\ \bibinfo {pages} {2122--2165}\BibitemShut {NoStop}%
\bibitem [{\citenamefont {Sauvage}\ and\ \citenamefont {Larocca}(2024)}]{sauvageClassicalShadowsSymmetries2024}%
  \BibitemOpen
  \bibfield  {author} {\bibinfo {author} {\bibfnamefont {F.}~\bibnamefont {Sauvage}}\ and\ \bibinfo {author} {\bibfnamefont {M.}~\bibnamefont {Larocca}},\ }\bibfield  {title} {\bibinfo {title} {Classical shadows with symmetries},\ }\href {https://doi.org/10.48550/arXiv.2408.05279} {\bibfield  {journal} {\bibinfo  {journal} {arXiv preprint arXiv:2408.05279}\ } (\bibinfo {year} {2024})}\BibitemShut {NoStop}%
\bibitem [{\citenamefont {Bu}\ \emph {et~al.}(2024)\citenamefont {Bu}, \citenamefont {Koh}, \citenamefont {Garcia},\ and\ \citenamefont {Jaffe}}]{buClassicalShadowsPauliinvariant2024}%
  \BibitemOpen
  \bibfield  {author} {\bibinfo {author} {\bibfnamefont {K.}~\bibnamefont {Bu}}, \bibinfo {author} {\bibfnamefont {D.~E.}\ \bibnamefont {Koh}}, \bibinfo {author} {\bibfnamefont {R.~J.}\ \bibnamefont {Garcia}},\ and\ \bibinfo {author} {\bibfnamefont {A.}~\bibnamefont {Jaffe}},\ }\bibfield  {title} {\bibinfo {title} {Classical shadows with {P}auli-invariant unitary ensembles},\ }\href {https://doi.org/10.1038/s41534-023-00801-w} {\bibfield  {journal} {\bibinfo  {journal} {npj Quantum Information}\ }\textbf {\bibinfo {volume} {10}},\ \bibinfo {pages} {6} (\bibinfo {year} {2024})}\BibitemShut {NoStop}%
\bibitem [{\citenamefont {Grier}\ \emph {et~al.}(2024{\natexlab{b}})\citenamefont {Grier}, \citenamefont {Pashayan},\ and\ \citenamefont {Schaeffer}}]{grierSampleoptimalClassicalShadows2024}%
  \BibitemOpen
  \bibfield  {author} {\bibinfo {author} {\bibfnamefont {D.}~\bibnamefont {Grier}}, \bibinfo {author} {\bibfnamefont {H.}~\bibnamefont {Pashayan}},\ and\ \bibinfo {author} {\bibfnamefont {L.}~\bibnamefont {Schaeffer}},\ }\bibfield  {title} {\bibinfo {title} {Sample-optimal classical shadows for pure states},\ }\href {https://doi.org/10.22331/q-2024-06-17-1373} {\bibfield  {journal} {\bibinfo  {journal} {Quantum}\ }\textbf {\bibinfo {volume} {8}},\ \bibinfo {pages} {1373} (\bibinfo {year} {2024}{\natexlab{b}})}\BibitemShut {NoStop}%
\bibitem [{\citenamefont {West}\ \emph {et~al.}(2024)\citenamefont {West}, \citenamefont {Mele}, \citenamefont {Larocca},\ and\ \citenamefont {Cerezo}}]{west2024real}%
  \BibitemOpen
  \bibfield  {author} {\bibinfo {author} {\bibfnamefont {M.}~\bibnamefont {West}}, \bibinfo {author} {\bibfnamefont {A.~A.}\ \bibnamefont {Mele}}, \bibinfo {author} {\bibfnamefont {M.}~\bibnamefont {Larocca}},\ and\ \bibinfo {author} {\bibfnamefont {M.}~\bibnamefont {Cerezo}},\ }\bibfield  {title} {\bibinfo {title} {Real classical shadows},\ }\href {https://doi.org/10.48550/arXiv.2410.23481} {\bibfield  {journal} {\bibinfo  {journal} {arXiv preprint arXiv:2410.23481}\ } (\bibinfo {year} {2024})}\BibitemShut {NoStop}%
\bibitem [{\citenamefont {Ippoliti}(2024)}]{ippoliti2024classical}%
  \BibitemOpen
  \bibfield  {author} {\bibinfo {author} {\bibfnamefont {M.}~\bibnamefont {Ippoliti}},\ }\bibfield  {title} {\bibinfo {title} {Classical shadows based on locally-entangled measurements},\ }\href {https://doi.org/10.22331/q-2024-03-21-1293} {\bibfield  {journal} {\bibinfo  {journal} {{Quantum}}\ }\textbf {\bibinfo {volume} {8}},\ \bibinfo {pages} {1293} (\bibinfo {year} {2024})}\BibitemShut {NoStop}%
\bibitem [{\citenamefont {King}\ \emph {et~al.}(2024)\citenamefont {King}, \citenamefont {Gosset}, \citenamefont {Kothari},\ and\ \citenamefont {Babbush}}]{king2024triply}%
  \BibitemOpen
  \bibfield  {author} {\bibinfo {author} {\bibfnamefont {R.}~\bibnamefont {King}}, \bibinfo {author} {\bibfnamefont {D.}~\bibnamefont {Gosset}}, \bibinfo {author} {\bibfnamefont {R.}~\bibnamefont {Kothari}},\ and\ \bibinfo {author} {\bibfnamefont {R.}~\bibnamefont {Babbush}},\ }\bibfield  {title} {\bibinfo {title} {Triply efficient shadow tomography},\ }\href {https://doi.org/10.48550/arXiv.2404.19211} {\bibfield  {journal} {\bibinfo  {journal} {arXiv preprint arXiv:2404.19211}\ } (\bibinfo {year} {2024})}\BibitemShut {NoStop}%
\bibitem [{\citenamefont {De~Palma}\ \emph {et~al.}(2024)\citenamefont {De~Palma}, \citenamefont {Klein},\ and\ \citenamefont {Pastorello}}]{de2024classical}%
  \BibitemOpen
  \bibfield  {author} {\bibinfo {author} {\bibfnamefont {G.}~\bibnamefont {De~Palma}}, \bibinfo {author} {\bibfnamefont {T.}~\bibnamefont {Klein}},\ and\ \bibinfo {author} {\bibfnamefont {D.}~\bibnamefont {Pastorello}},\ }\bibfield  {title} {\bibinfo {title} {Classical shadows meet quantum optimal mass transport},\ }\href {https://doi.org/10.1063/5.0178897} {\bibfield  {journal} {\bibinfo  {journal} {Journal of Mathematical Physics}\ }\textbf {\bibinfo {volume} {65}},\ \bibinfo {pages} {092201} (\bibinfo {year} {2024})}\BibitemShut {NoStop}%
\bibitem [{\citenamefont {Becker}\ \emph {et~al.}(2024)\citenamefont {Becker}, \citenamefont {Datta}, \citenamefont {Lami},\ and\ \citenamefont {Rouze}}]{becker2024classical}%
  \BibitemOpen
  \bibfield  {author} {\bibinfo {author} {\bibfnamefont {S.}~\bibnamefont {Becker}}, \bibinfo {author} {\bibfnamefont {N.}~\bibnamefont {Datta}}, \bibinfo {author} {\bibfnamefont {L.}~\bibnamefont {Lami}},\ and\ \bibinfo {author} {\bibfnamefont {C.}~\bibnamefont {Rouze}},\ }\bibfield  {title} {\bibinfo {title} {Classical shadow tomography for continuous variables quantum systems},\ }\href {https://doi.org/10.1109/TIT.2024.3357972} {\bibfield  {journal} {\bibinfo  {journal} {IEEE Transactions on Information Theory}\ }\textbf {\bibinfo {volume} {70}},\ \bibinfo {pages} {3427} (\bibinfo {year} {2024})}\BibitemShut {NoStop}%
\bibitem [{\citenamefont {Hearth}\ \emph {et~al.}(2024)\citenamefont {Hearth}, \citenamefont {Flynn}, \citenamefont {Chandran},\ and\ \citenamefont {Laumann}}]{hearth2024efficient}%
  \BibitemOpen
  \bibfield  {author} {\bibinfo {author} {\bibfnamefont {S.~N.}\ \bibnamefont {Hearth}}, \bibinfo {author} {\bibfnamefont {M.~O.}\ \bibnamefont {Flynn}}, \bibinfo {author} {\bibfnamefont {A.}~\bibnamefont {Chandran}},\ and\ \bibinfo {author} {\bibfnamefont {C.~R.}\ \bibnamefont {Laumann}},\ }\bibfield  {title} {\bibinfo {title} {Efficient local classical shadow tomography with number conservation},\ }\href {https://doi.org/10.1103/PhysRevLett.133.060802} {\bibfield  {journal} {\bibinfo  {journal} {Phys. Rev. Lett.}\ }\textbf {\bibinfo {volume} {133}},\ \bibinfo {pages} {060802} (\bibinfo {year} {2024})}\BibitemShut {NoStop}%
\bibitem [{\citenamefont {Cai}\ \emph {et~al.}(2024)\citenamefont {Cai}, \citenamefont {Chapman}, \citenamefont {Jnane},\ and\ \citenamefont {Koczor}}]{cai2024biased}%
  \BibitemOpen
  \bibfield  {author} {\bibinfo {author} {\bibfnamefont {Z.}~\bibnamefont {Cai}}, \bibinfo {author} {\bibfnamefont {A.}~\bibnamefont {Chapman}}, \bibinfo {author} {\bibfnamefont {H.}~\bibnamefont {Jnane}},\ and\ \bibinfo {author} {\bibfnamefont {B.}~\bibnamefont {Koczor}},\ }\bibfield  {title} {\bibinfo {title} {Biased estimator channels for classical shadows},\ }\href {https://doi.org/10.48550/arXiv.2402.09511} {\bibfield  {journal} {\bibinfo  {journal} {arXiv preprint arXiv:2402.09511}\ } (\bibinfo {year} {2024})}\BibitemShut {NoStop}%
\bibitem [{\citenamefont {Chen}\ \emph {et~al.}(2024)\citenamefont {Chen}, \citenamefont {Saleem},\ and\ \citenamefont {Perlin}}]{chen2024quantum}%
  \BibitemOpen
  \bibfield  {author} {\bibinfo {author} {\bibfnamefont {D.~T.~S.}\ \bibnamefont {Chen}}, \bibinfo {author} {\bibfnamefont {Z.~H.}\ \bibnamefont {Saleem}},\ and\ \bibinfo {author} {\bibfnamefont {M.~A.}\ \bibnamefont {Perlin}},\ }\bibfield  {title} {\bibinfo {title} {Quantum circuit cutting for classical shadows},\ }\href {https://doi.org/10.1145/3665335} {\bibfield  {journal} {\bibinfo  {journal} {ACM Transactions on Quantum Computing}\ }\textbf {\bibinfo {volume} {5(2)}},\ \bibinfo {pages} {13} (\bibinfo {year} {2024})}\BibitemShut {NoStop}%
\bibitem [{\citenamefont {Heyraud}\ \emph {et~al.}(2024)\citenamefont {Heyraud}, \citenamefont {Chomet},\ and\ \citenamefont {Tilly}}]{heyraud2024unified}%
  \BibitemOpen
  \bibfield  {author} {\bibinfo {author} {\bibfnamefont {V.}~\bibnamefont {Heyraud}}, \bibinfo {author} {\bibfnamefont {H.}~\bibnamefont {Chomet}},\ and\ \bibinfo {author} {\bibfnamefont {J.}~\bibnamefont {Tilly}},\ }\bibfield  {title} {\bibinfo {title} {Unified framework for matchgate classical shadows},\ }\href {https://doi.org/10.48550/arXiv.2409.03836} {\bibfield  {journal} {\bibinfo  {journal} {arXiv preprint arXiv:2409.03836}\ } (\bibinfo {year} {2024})}\BibitemShut {NoStop}%
\bibitem [{\citenamefont {Akhtar}\ \emph {et~al.}(2024)\citenamefont {Akhtar}, \citenamefont {Anand}, \citenamefont {Marshall},\ and\ \citenamefont {You}}]{akhtar2024dual}%
  \BibitemOpen
  \bibfield  {author} {\bibinfo {author} {\bibfnamefont {A.~A.}\ \bibnamefont {Akhtar}}, \bibinfo {author} {\bibfnamefont {N.}~\bibnamefont {Anand}}, \bibinfo {author} {\bibfnamefont {J.}~\bibnamefont {Marshall}},\ and\ \bibinfo {author} {\bibfnamefont {Y.-Z.}\ \bibnamefont {You}},\ }\bibfield  {title} {\bibinfo {title} {Dual-unitary classical shadow tomography},\ }\href {https://doi.org/10.48550/arXiv.2404.01068} {\bibfield  {journal} {\bibinfo  {journal} {arXiv preprint arXiv:2404.01068}\ } (\bibinfo {year} {2024})}\BibitemShut {NoStop}%
\bibitem [{\citenamefont {Wang}(2024)}]{wang2024quantum}%
  \BibitemOpen
  \bibfield  {author} {\bibinfo {author} {\bibfnamefont {Y.}~\bibnamefont {Wang}},\ }\bibfield  {title} {\bibinfo {title} {Quantum advantage via efficient post-processing on qudit shadow tomography},\ }\href {https://doi.org/10.48550/arXiv.2408.16244} {\bibfield  {journal} {\bibinfo  {journal} {arXiv preprint arXiv:2408.16244}\ } (\bibinfo {year} {2024})}\BibitemShut {NoStop}%
\bibitem [{\citenamefont {Wu}\ \emph {et~al.}(2024)\citenamefont {Wu}, \citenamefont {Wang}, \citenamefont {Yao}, \citenamefont {Zhai}, \citenamefont {You},\ and\ \citenamefont {Zhang}}]{wu2024contractive}%
  \BibitemOpen
  \bibfield  {author} {\bibinfo {author} {\bibfnamefont {Y.}~\bibnamefont {Wu}}, \bibinfo {author} {\bibfnamefont {C.}~\bibnamefont {Wang}}, \bibinfo {author} {\bibfnamefont {J.}~\bibnamefont {Yao}}, \bibinfo {author} {\bibfnamefont {H.}~\bibnamefont {Zhai}}, \bibinfo {author} {\bibfnamefont {Y.-Z.}\ \bibnamefont {You}},\ and\ \bibinfo {author} {\bibfnamefont {P.}~\bibnamefont {Zhang}},\ }\bibfield  {title} {\bibinfo {title} {Contractive unitary and classical shadow tomography},\ }\href {https://doi.org/10.48550/arXiv.2412.01850} {\bibfield  {journal} {\bibinfo  {journal} {arXiv preprint arXiv:2412.01850}\ } (\bibinfo {year} {2024})}\BibitemShut {NoStop}%
\bibitem [{\citenamefont {Zhou}\ and\ \citenamefont {Zhang}(2024)}]{zhou2024efficient}%
  \BibitemOpen
  \bibfield  {author} {\bibinfo {author} {\bibfnamefont {T.-G.}\ \bibnamefont {Zhou}}\ and\ \bibinfo {author} {\bibfnamefont {P.}~\bibnamefont {Zhang}},\ }\bibfield  {title} {\bibinfo {title} {Efficient {C}lassical {S}hadow {T}omography through {M}any-body {L}ocalization {D}ynamics},\ }\href {https://doi.org/10.22331/q-2024-09-11-1467} {\bibfield  {journal} {\bibinfo  {journal} {{Quantum}}\ }\textbf {\bibinfo {volume} {8}},\ \bibinfo {pages} {1467} (\bibinfo {year} {2024})}\BibitemShut {NoStop}%
\bibitem [{\citenamefont {Cioli}\ \emph {et~al.}(2024)\citenamefont {Cioli}, \citenamefont {Ercolessi}, \citenamefont {Ippoliti}, \citenamefont {Turkeshi},\ and\ \citenamefont {Piroli}}]{cioli2024approximate}%
  \BibitemOpen
  \bibfield  {author} {\bibinfo {author} {\bibfnamefont {R.}~\bibnamefont {Cioli}}, \bibinfo {author} {\bibfnamefont {E.}~\bibnamefont {Ercolessi}}, \bibinfo {author} {\bibfnamefont {M.}~\bibnamefont {Ippoliti}}, \bibinfo {author} {\bibfnamefont {X.}~\bibnamefont {Turkeshi}},\ and\ \bibinfo {author} {\bibfnamefont {L.}~\bibnamefont {Piroli}},\ }\bibfield  {title} {\bibinfo {title} {Approximate inverse measurement channel for shallow shadows},\ }\href {https://doi.org/10.48550/arXiv.2407.11813} {\bibfield  {journal} {\bibinfo  {journal} {arXiv preprint arXiv:2407.11813}\ } (\bibinfo {year} {2024})}\BibitemShut {NoStop}%
\bibitem [{\citenamefont {Liu}\ \emph {et~al.}(2024)\citenamefont {Liu}, \citenamefont {Li}, \citenamefont {Yuan}, \citenamefont {Zhu},\ and\ \citenamefont {Zhou}}]{liu2024auxiliary}%
  \BibitemOpen
  \bibfield  {author} {\bibinfo {author} {\bibfnamefont {Q.}~\bibnamefont {Liu}}, \bibinfo {author} {\bibfnamefont {Z.}~\bibnamefont {Li}}, \bibinfo {author} {\bibfnamefont {X.}~\bibnamefont {Yuan}}, \bibinfo {author} {\bibfnamefont {H.}~\bibnamefont {Zhu}},\ and\ \bibinfo {author} {\bibfnamefont {Y.}~\bibnamefont {Zhou}},\ }\bibfield  {title} {\bibinfo {title} {Auxiliary-free replica shadow estimation},\ }\href {https://doi.org/10.48550/arXiv.2407.20865} {\bibfield  {journal} {\bibinfo  {journal} {arXiv preprint arXiv:2407.20865}\ } (\bibinfo {year} {2024})}\BibitemShut {NoStop}%
\bibitem [{\citenamefont {Bertoni}\ \emph {et~al.}(2024)\citenamefont {Bertoni}, \citenamefont {Haferkamp}, \citenamefont {Hinsche}, \citenamefont {Ioannou}, \citenamefont {Eisert},\ and\ \citenamefont {Pashayan}}]{bertoni2024shallow}%
  \BibitemOpen
  \bibfield  {author} {\bibinfo {author} {\bibfnamefont {C.}~\bibnamefont {Bertoni}}, \bibinfo {author} {\bibfnamefont {J.}~\bibnamefont {Haferkamp}}, \bibinfo {author} {\bibfnamefont {M.}~\bibnamefont {Hinsche}}, \bibinfo {author} {\bibfnamefont {M.}~\bibnamefont {Ioannou}}, \bibinfo {author} {\bibfnamefont {J.}~\bibnamefont {Eisert}},\ and\ \bibinfo {author} {\bibfnamefont {H.}~\bibnamefont {Pashayan}},\ }\bibfield  {title} {\bibinfo {title} {Shallow shadows: Expectation estimation using low-depth random {C}lifford circuits},\ }\href {https://doi.org/10.1103/PhysRevLett.133.020602} {\bibfield  {journal} {\bibinfo  {journal} {Phys. Rev. Lett.}\ }\textbf {\bibinfo {volume} {133}},\ \bibinfo {pages} {020602} (\bibinfo {year} {2024})}\BibitemShut {NoStop}%
\bibitem [{\citenamefont {Anselmetti}\ \emph {et~al.}(2024)\citenamefont {Anselmetti}, \citenamefont {Degroote}, \citenamefont {Moll}, \citenamefont {Santagati},\ and\ \citenamefont {Streif}}]{anselmetti2024classical}%
  \BibitemOpen
  \bibfield  {author} {\bibinfo {author} {\bibfnamefont {G.-L.~R.}\ \bibnamefont {Anselmetti}}, \bibinfo {author} {\bibfnamefont {M.}~\bibnamefont {Degroote}}, \bibinfo {author} {\bibfnamefont {N.}~\bibnamefont {Moll}}, \bibinfo {author} {\bibfnamefont {R.}~\bibnamefont {Santagati}},\ and\ \bibinfo {author} {\bibfnamefont {M.}~\bibnamefont {Streif}},\ }\bibfield  {title} {\bibinfo {title} {Classical optimisation of reduced density matrix estimations with classical shadows using {N}-representability conditions under shot noise considerations},\ }\href {https://doi.org/10.48550/arXiv.2411.18430} {\bibfield  {journal} {\bibinfo  {journal} {arXiv preprint arXiv:2411.18430}\ } (\bibinfo {year} {2024})}\BibitemShut {NoStop}%
\bibitem [{\citenamefont {Mao}\ \emph {et~al.}(2024)\citenamefont {Mao}, \citenamefont {Yi},\ and\ \citenamefont {Zhu}}]{mao2024magic}%
  \BibitemOpen
  \bibfield  {author} {\bibinfo {author} {\bibfnamefont {C.}~\bibnamefont {Mao}}, \bibinfo {author} {\bibfnamefont {C.}~\bibnamefont {Yi}},\ and\ \bibinfo {author} {\bibfnamefont {H.}~\bibnamefont {Zhu}},\ }\bibfield  {title} {\bibinfo {title} {The magic in qudit shadow estimation based on the {C}lifford group},\ }\href {https://doi.org/10.48550/arXiv.2410.13572} {\bibfield  {journal} {\bibinfo  {journal} {arXiv preprint arXiv:2410.13572}\ } (\bibinfo {year} {2024})}\BibitemShut {NoStop}%
\bibitem [{\citenamefont {Caprotti}\ \emph {et~al.}(2024)\citenamefont {Caprotti}, \citenamefont {Morris},\ and\ \citenamefont {Daki\ifmmode~\acute{c}\else \'{c}\fi{}}}]{caprotti2024optimizing}%
  \BibitemOpen
  \bibfield  {author} {\bibinfo {author} {\bibfnamefont {A.}~\bibnamefont {Caprotti}}, \bibinfo {author} {\bibfnamefont {J.}~\bibnamefont {Morris}},\ and\ \bibinfo {author} {\bibfnamefont {B.}~\bibnamefont {Daki\ifmmode~\acute{c}\else \'{c}\fi{}}},\ }\bibfield  {title} {\bibinfo {title} {Optimizing quantum tomography via shadow inversion},\ }\href {https://doi.org/10.1103/PhysRevResearch.6.033301} {\bibfield  {journal} {\bibinfo  {journal} {Phys. Rev. Res.}\ }\textbf {\bibinfo {volume} {6}},\ \bibinfo {pages} {033301} (\bibinfo {year} {2024})}\BibitemShut {NoStop}%
\bibitem [{\citenamefont {Zhang}\ \emph {et~al.}(2023)\citenamefont {Zhang}, \citenamefont {Nakaji}, \citenamefont {Choi},\ and\ \citenamefont {Aspuru-Guzik}}]{zhang2023composite}%
  \BibitemOpen
  \bibfield  {author} {\bibinfo {author} {\bibfnamefont {Z.-J.}\ \bibnamefont {Zhang}}, \bibinfo {author} {\bibfnamefont {K.}~\bibnamefont {Nakaji}}, \bibinfo {author} {\bibfnamefont {M.}~\bibnamefont {Choi}},\ and\ \bibinfo {author} {\bibfnamefont {A.}~\bibnamefont {Aspuru-Guzik}},\ }\bibfield  {title} {\bibinfo {title} {A composite measurement scheme for efficient quantum observable estimation},\ }\href {https://doi.org/10.48550/arXiv.2305.02439} {\bibfield  {journal} {\bibinfo  {journal} {arXiv preprint arXiv:2305.02439}\ } (\bibinfo {year} {2023})}\BibitemShut {NoStop}%
\bibitem [{\citenamefont {Huang}\ \emph {et~al.}(2024)\citenamefont {Huang}, \citenamefont {Chen}, \citenamefont {Gupt}, \citenamefont {Suchara}, \citenamefont {Tran}, \citenamefont {McArdle},\ and\ \citenamefont {Galli}}]{huang2024evaluating}%
  \BibitemOpen
  \bibfield  {author} {\bibinfo {author} {\bibfnamefont {B.}~\bibnamefont {Huang}}, \bibinfo {author} {\bibfnamefont {Y.-T.}\ \bibnamefont {Chen}}, \bibinfo {author} {\bibfnamefont {B.}~\bibnamefont {Gupt}}, \bibinfo {author} {\bibfnamefont {M.}~\bibnamefont {Suchara}}, \bibinfo {author} {\bibfnamefont {A.}~\bibnamefont {Tran}}, \bibinfo {author} {\bibfnamefont {S.}~\bibnamefont {McArdle}},\ and\ \bibinfo {author} {\bibfnamefont {G.}~\bibnamefont {Galli}},\ }\bibfield  {title} {\bibinfo {title} {Evaluating a quantum-classical quantum {Monte Carlo} algorithm with matchgate shadows},\ }\href {https://doi.org/10.1103/PhysRevResearch.6.043063} {\bibfield  {journal} {\bibinfo  {journal} {Phys. Rev. Res.}\ }\textbf {\bibinfo {volume} {6}},\ \bibinfo {pages} {043063} (\bibinfo {year} {2024})}\BibitemShut {NoStop}%
\bibitem [{\citenamefont {Avdic}\ and\ \citenamefont {Mazziotti}(2024{\natexlab{a}})}]{avdic2024enhanced}%
  \BibitemOpen
  \bibfield  {author} {\bibinfo {author} {\bibfnamefont {I.}~\bibnamefont {Avdic}}\ and\ \bibinfo {author} {\bibfnamefont {D.~A.}\ \bibnamefont {Mazziotti}},\ }\bibfield  {title} {\bibinfo {title} {Enhanced shadow tomography of molecular excited states via the enforcement of {$N$}-representability conditions by semidefinite programming},\ }\href {https://doi.org/10.1103/PhysRevA.110.052407} {\bibfield  {journal} {\bibinfo  {journal} {Phys. Rev. A}\ }\textbf {\bibinfo {volume} {110}},\ \bibinfo {pages} {052407} (\bibinfo {year} {2024}{\natexlab{a}})}\BibitemShut {NoStop}%
\bibitem [{\citenamefont {Wang}\ \emph {et~al.}(2024)\citenamefont {Wang}, \citenamefont {Avdic},\ and\ \citenamefont {Mazziotti}}]{wang2024shadow}%
  \BibitemOpen
  \bibfield  {author} {\bibinfo {author} {\bibfnamefont {Y.}~\bibnamefont {Wang}}, \bibinfo {author} {\bibfnamefont {I.}~\bibnamefont {Avdic}},\ and\ \bibinfo {author} {\bibfnamefont {D.~A.}\ \bibnamefont {Mazziotti}},\ }\bibfield  {title} {\bibinfo {title} {Shadow ansatz for the many-fermion wave function in scalable molecular simulations on quantum computing devices},\ }\href {https://doi.org/10.48550/arXiv.2408.11026} {\bibfield  {journal} {\bibinfo  {journal} {arXiv preprint arXiv:2408.11026}\ } (\bibinfo {year} {2024})}\BibitemShut {NoStop}%
\bibitem [{\citenamefont {Avdic}\ and\ \citenamefont {Mazziotti}(2024{\natexlab{b}})}]{avdic2024fewer}%
  \BibitemOpen
  \bibfield  {author} {\bibinfo {author} {\bibfnamefont {I.}~\bibnamefont {Avdic}}\ and\ \bibinfo {author} {\bibfnamefont {D.~A.}\ \bibnamefont {Mazziotti}},\ }\bibfield  {title} {\bibinfo {title} {Fewer measurements from shadow tomography with {$N$}-representability conditions},\ }\href {https://doi.org/10.1103/PhysRevLett.132.220802} {\bibfield  {journal} {\bibinfo  {journal} {Phys. Rev. Lett.}\ }\textbf {\bibinfo {volume} {132}},\ \bibinfo {pages} {220802} (\bibinfo {year} {2024}{\natexlab{b}})}\BibitemShut {NoStop}%
\bibitem [{\citenamefont {Huang}\ \emph {et~al.}(2022)\citenamefont {Huang}, \citenamefont {Kueng}, \citenamefont {Torlai}, \citenamefont {Albert},\ and\ \citenamefont {Preskill}}]{huangProvablyEfficientMachine2022}%
  \BibitemOpen
  \bibfield  {author} {\bibinfo {author} {\bibfnamefont {H.-Y.}\ \bibnamefont {Huang}}, \bibinfo {author} {\bibfnamefont {R.}~\bibnamefont {Kueng}}, \bibinfo {author} {\bibfnamefont {G.}~\bibnamefont {Torlai}}, \bibinfo {author} {\bibfnamefont {V.~V.}\ \bibnamefont {Albert}},\ and\ \bibinfo {author} {\bibfnamefont {J.}~\bibnamefont {Preskill}},\ }\bibfield  {title} {\bibinfo {title} {Provably efficient machine learning for quantum many-body problems},\ }\href {https://doi.org/10.1126/science.abk3333} {\bibfield  {journal} {\bibinfo  {journal} {Science}\ }\textbf {\bibinfo {volume} {377}},\ \bibinfo {pages} {eabk3333} (\bibinfo {year} {2022})}\BibitemShut {NoStop}%
\bibitem [{\citenamefont {Lewis}\ \emph {et~al.}(2024)\citenamefont {Lewis}, \citenamefont {Huang}, \citenamefont {Tran}, \citenamefont {Lehner}, \citenamefont {Kueng},\ and\ \citenamefont {Preskill}}]{lewisImprovedMachineLearning2024}%
  \BibitemOpen
  \bibfield  {author} {\bibinfo {author} {\bibfnamefont {L.}~\bibnamefont {Lewis}}, \bibinfo {author} {\bibfnamefont {H.-Y.}\ \bibnamefont {Huang}}, \bibinfo {author} {\bibfnamefont {V.~T.}\ \bibnamefont {Tran}}, \bibinfo {author} {\bibfnamefont {S.}~\bibnamefont {Lehner}}, \bibinfo {author} {\bibfnamefont {R.}~\bibnamefont {Kueng}},\ and\ \bibinfo {author} {\bibfnamefont {J.}~\bibnamefont {Preskill}},\ }\bibfield  {title} {\bibinfo {title} {Improved machine learning algorithm for predicting ground state properties},\ }\href {https://doi.org/10.1038/s41467-024-45014-7} {\bibfield  {journal} {\bibinfo  {journal} {Nature Communications}\ }\textbf {\bibinfo {volume} {15}},\ \bibinfo {pages} {895} (\bibinfo {year} {2024})}\BibitemShut {NoStop}%
\bibitem [{\citenamefont {Jerbi}\ \emph {et~al.}(2024)\citenamefont {Jerbi}, \citenamefont {Gyurik}, \citenamefont {Marshall}, \citenamefont {Molteni},\ and\ \citenamefont {Dunjko}}]{jerbiShadowsQuantumMachine2024}%
  \BibitemOpen
  \bibfield  {author} {\bibinfo {author} {\bibfnamefont {S.}~\bibnamefont {Jerbi}}, \bibinfo {author} {\bibfnamefont {C.}~\bibnamefont {Gyurik}}, \bibinfo {author} {\bibfnamefont {S.~C.}\ \bibnamefont {Marshall}}, \bibinfo {author} {\bibfnamefont {R.}~\bibnamefont {Molteni}},\ and\ \bibinfo {author} {\bibfnamefont {V.}~\bibnamefont {Dunjko}},\ }\bibfield  {title} {\bibinfo {title} {Shadows of quantum machine learning},\ }\href {https://doi.org/10.1038/s41467-024-49877-8} {\bibfield  {journal} {\bibinfo  {journal} {Nature Communications}\ }\textbf {\bibinfo {volume} {15}},\ \bibinfo {pages} {5676} (\bibinfo {year} {2024})}\BibitemShut {NoStop}%
\bibitem [{\citenamefont {Zhao}\ and\ \citenamefont {Miyake}(2024)}]{zhaoGrouptheoreticErrorMitigation2024}%
  \BibitemOpen
  \bibfield  {author} {\bibinfo {author} {\bibfnamefont {A.}~\bibnamefont {Zhao}}\ and\ \bibinfo {author} {\bibfnamefont {A.}~\bibnamefont {Miyake}},\ }\bibfield  {title} {\bibinfo {title} {Group-theoretic error mitigation enabled by classical shadows and symmetries},\ }\href {https://doi.org/10.1038/s41534-024-00854-5} {\bibfield  {journal} {\bibinfo  {journal} {npj Quantum Information}\ }\textbf {\bibinfo {volume} {10}},\ \bibinfo {pages} {57} (\bibinfo {year} {2024})}\BibitemShut {NoStop}%
\bibitem [{\citenamefont {Seif}\ \emph {et~al.}(2023)\citenamefont {Seif}, \citenamefont {Cian}, \citenamefont {Zhou}, \citenamefont {Chen},\ and\ \citenamefont {Jiang}}]{seifShadowDistillationQuantum2023}%
  \BibitemOpen
  \bibfield  {author} {\bibinfo {author} {\bibfnamefont {A.}~\bibnamefont {Seif}}, \bibinfo {author} {\bibfnamefont {Z.-P.}\ \bibnamefont {Cian}}, \bibinfo {author} {\bibfnamefont {S.}~\bibnamefont {Zhou}}, \bibinfo {author} {\bibfnamefont {S.}~\bibnamefont {Chen}},\ and\ \bibinfo {author} {\bibfnamefont {L.}~\bibnamefont {Jiang}},\ }\bibfield  {title} {\bibinfo {title} {Shadow {{Distillation}}: {{Quantum Error Mitigation}} with {{Classical Shadows}} for {{Near-Term Quantum Processors}}},\ }\href {https://doi.org/10.1103/PRXQuantum.4.010303} {\bibfield  {journal} {\bibinfo  {journal} {PRX Quantum}\ }\textbf {\bibinfo {volume} {4}},\ \bibinfo {pages} {010303} (\bibinfo {year} {2023})}\BibitemShut {NoStop}%
\bibitem [{\citenamefont {Elben}\ \emph {et~al.}(2020)\citenamefont {Elben}, \citenamefont {Kueng}, \citenamefont {Huang}, \citenamefont {van Bijnen}, \citenamefont {Kokail}, \citenamefont {Dalmonte}, \citenamefont {Calabrese}, \citenamefont {Kraus}, \citenamefont {Preskill}, \citenamefont {Zoller},\ and\ \citenamefont {Vermersch}}]{elban2020mixed}%
  \BibitemOpen
  \bibfield  {author} {\bibinfo {author} {\bibfnamefont {A.}~\bibnamefont {Elben}}, \bibinfo {author} {\bibfnamefont {R.}~\bibnamefont {Kueng}}, \bibinfo {author} {\bibfnamefont {H.-Y.~R.}\ \bibnamefont {Huang}}, \bibinfo {author} {\bibfnamefont {R.}~\bibnamefont {van Bijnen}}, \bibinfo {author} {\bibfnamefont {C.}~\bibnamefont {Kokail}}, \bibinfo {author} {\bibfnamefont {M.}~\bibnamefont {Dalmonte}}, \bibinfo {author} {\bibfnamefont {P.}~\bibnamefont {Calabrese}}, \bibinfo {author} {\bibfnamefont {B.}~\bibnamefont {Kraus}}, \bibinfo {author} {\bibfnamefont {J.}~\bibnamefont {Preskill}}, \bibinfo {author} {\bibfnamefont {P.}~\bibnamefont {Zoller}},\ and\ \bibinfo {author} {\bibfnamefont {B.}~\bibnamefont {Vermersch}},\ }\bibfield  {title} {\bibinfo {title} {Mixed-state entanglement from local randomized measurements},\ }\href {https://doi.org/10.1103/PhysRevLett.125.200501} {\bibfield  {journal} {\bibinfo  {journal} {Phys. Rev. Lett.}\ }\textbf {\bibinfo {volume} {125}},\ \bibinfo {pages} {200501} (\bibinfo
  {year} {2020})}\BibitemShut {NoStop}%
\bibitem [{\citenamefont {Neven}\ \emph {et~al.}(2021)\citenamefont {Neven}, \citenamefont {Carrasco}, \citenamefont {Vitale}, \citenamefont {Kokail}, \citenamefont {Elben}, \citenamefont {Dalmonte}, \citenamefont {Calabrese}, \citenamefont {Zoller}, \citenamefont {Vermersch}, \citenamefont {Kueng} \emph {et~al.}}]{neven2021symmetry}%
  \BibitemOpen
  \bibfield  {author} {\bibinfo {author} {\bibfnamefont {A.}~\bibnamefont {Neven}}, \bibinfo {author} {\bibfnamefont {J.}~\bibnamefont {Carrasco}}, \bibinfo {author} {\bibfnamefont {V.}~\bibnamefont {Vitale}}, \bibinfo {author} {\bibfnamefont {C.}~\bibnamefont {Kokail}}, \bibinfo {author} {\bibfnamefont {A.}~\bibnamefont {Elben}}, \bibinfo {author} {\bibfnamefont {M.}~\bibnamefont {Dalmonte}}, \bibinfo {author} {\bibfnamefont {P.}~\bibnamefont {Calabrese}}, \bibinfo {author} {\bibfnamefont {P.}~\bibnamefont {Zoller}}, \bibinfo {author} {\bibfnamefont {B.}~\bibnamefont {Vermersch}}, \bibinfo {author} {\bibfnamefont {R.}~\bibnamefont {Kueng}}, \emph {et~al.},\ }\bibfield  {title} {\bibinfo {title} {Symmetry-resolved entanglement detection using partial transpose moments},\ }\href {https://doi.org/10.1038/s41534-021-00487-y} {\bibfield  {journal} {\bibinfo  {journal} {npj Quantum Information}\ }\textbf {\bibinfo {volume} {7}},\ \bibinfo {pages} {152} (\bibinfo {year} {2021})}\BibitemShut {NoStop}%
\bibitem [{\citenamefont {Rath}\ \emph {et~al.}(2021)\citenamefont {Rath}, \citenamefont {Branciard}, \citenamefont {Minguzzi},\ and\ \citenamefont {Vermersch}}]{rath2021quantum}%
  \BibitemOpen
  \bibfield  {author} {\bibinfo {author} {\bibfnamefont {A.}~\bibnamefont {Rath}}, \bibinfo {author} {\bibfnamefont {C.}~\bibnamefont {Branciard}}, \bibinfo {author} {\bibfnamefont {A.}~\bibnamefont {Minguzzi}},\ and\ \bibinfo {author} {\bibfnamefont {B.}~\bibnamefont {Vermersch}},\ }\bibfield  {title} {\bibinfo {title} {Quantum {F}isher information from randomized measurements},\ }\href {https://doi.org/10.1103/PhysRevLett.127.260501} {\bibfield  {journal} {\bibinfo  {journal} {Phys. Rev. Lett.}\ }\textbf {\bibinfo {volume} {127}},\ \bibinfo {pages} {260501} (\bibinfo {year} {2021})}\BibitemShut {NoStop}%
\bibitem [{\citenamefont {Boyd}\ and\ \citenamefont {Koczor}(2022)}]{boyd2022training}%
  \BibitemOpen
  \bibfield  {author} {\bibinfo {author} {\bibfnamefont {G.}~\bibnamefont {Boyd}}\ and\ \bibinfo {author} {\bibfnamefont {B.}~\bibnamefont {Koczor}},\ }\bibfield  {title} {\bibinfo {title} {Training variational quantum circuits with {CoVaR}: Covariance root finding with classical shadows},\ }\href {https://doi.org/10.1103/PhysRevX.12.041022} {\bibfield  {journal} {\bibinfo  {journal} {Phys. Rev. X}\ }\textbf {\bibinfo {volume} {12}},\ \bibinfo {pages} {041022} (\bibinfo {year} {2022})}\BibitemShut {NoStop}%
\bibitem [{\citenamefont {Sack}\ \emph {et~al.}(2022)\citenamefont {Sack}, \citenamefont {Medina}, \citenamefont {Michailidis}, \citenamefont {Kueng},\ and\ \citenamefont {Serbyn}}]{sackAvoidingBarrenPlateaus2022}%
  \BibitemOpen
  \bibfield  {author} {\bibinfo {author} {\bibfnamefont {S.~H.}\ \bibnamefont {Sack}}, \bibinfo {author} {\bibfnamefont {R.~A.}\ \bibnamefont {Medina}}, \bibinfo {author} {\bibfnamefont {A.~A.}\ \bibnamefont {Michailidis}}, \bibinfo {author} {\bibfnamefont {R.}~\bibnamefont {Kueng}},\ and\ \bibinfo {author} {\bibfnamefont {M.}~\bibnamefont {Serbyn}},\ }\bibfield  {title} {\bibinfo {title} {Avoiding {{Barren Plateaus Using Classical Shadows}}},\ }\href {https://doi.org/10.1103/PRXQuantum.3.020365} {\bibfield  {journal} {\bibinfo  {journal} {PRX Quantum}\ }\textbf {\bibinfo {volume} {3}},\ \bibinfo {pages} {020365} (\bibinfo {year} {2022})}\BibitemShut {NoStop}%
\bibitem [{\citenamefont {Basheer}\ \emph {et~al.}(2023)\citenamefont {Basheer}, \citenamefont {Feng}, \citenamefont {Ferrie},\ and\ \citenamefont {Li}}]{basheer2023alternating}%
  \BibitemOpen
  \bibfield  {author} {\bibinfo {author} {\bibfnamefont {A.}~\bibnamefont {Basheer}}, \bibinfo {author} {\bibfnamefont {Y.}~\bibnamefont {Feng}}, \bibinfo {author} {\bibfnamefont {C.}~\bibnamefont {Ferrie}},\ and\ \bibinfo {author} {\bibfnamefont {S.}~\bibnamefont {Li}},\ }\bibfield  {title} {\bibinfo {title} {Alternating layered variational quantum circuits can be classically optimized efficiently using classical shadows},\ }in\ \href {https://doi.org/10.1609/aaai.v37i6.25830} {\emph {\bibinfo {booktitle} {Proceedings of the AAAI Conference on Artificial Intelligence}}},\ Vol.~\bibinfo {volume} {37}\ (\bibinfo {year} {2023})\ pp.\ \bibinfo {pages} {6770--6778}\BibitemShut {NoStop}%
\bibitem [{\citenamefont {Nakaji}\ \emph {et~al.}(2023)\citenamefont {Nakaji}, \citenamefont {Endo}, \citenamefont {Matsuzaki},\ and\ \citenamefont {Hakoshima}}]{nakaji2023measurement}%
  \BibitemOpen
  \bibfield  {author} {\bibinfo {author} {\bibfnamefont {K.}~\bibnamefont {Nakaji}}, \bibinfo {author} {\bibfnamefont {S.}~\bibnamefont {Endo}}, \bibinfo {author} {\bibfnamefont {Y.}~\bibnamefont {Matsuzaki}},\ and\ \bibinfo {author} {\bibfnamefont {H.}~\bibnamefont {Hakoshima}},\ }\bibfield  {title} {\bibinfo {title} {Measurement optimization of variational quantum simulation by classical shadow and derandomization},\ }\href {https://doi.org/10.22331/q-2023-05-04-995} {\bibfield  {journal} {\bibinfo  {journal} {{Quantum}}\ }\textbf {\bibinfo {volume} {7}},\ \bibinfo {pages} {995} (\bibinfo {year} {2023})}\BibitemShut {NoStop}%
\bibitem [{\citenamefont {Garcia}\ \emph {et~al.}(2021)\citenamefont {Garcia}, \citenamefont {Zhou},\ and\ \citenamefont {Jaffe}}]{garcia2021quantum}%
  \BibitemOpen
  \bibfield  {author} {\bibinfo {author} {\bibfnamefont {R.~J.}\ \bibnamefont {Garcia}}, \bibinfo {author} {\bibfnamefont {Y.}~\bibnamefont {Zhou}},\ and\ \bibinfo {author} {\bibfnamefont {A.}~\bibnamefont {Jaffe}},\ }\bibfield  {title} {\bibinfo {title} {Quantum scrambling with classical shadows},\ }\href {https://doi.org/10.1103/PhysRevResearch.3.033155} {\bibfield  {journal} {\bibinfo  {journal} {Phys. Rev. Res.}\ }\textbf {\bibinfo {volume} {3}},\ \bibinfo {pages} {033155} (\bibinfo {year} {2021})}\BibitemShut {NoStop}%
\bibitem [{\citenamefont {Helsen}\ \emph {et~al.}(2023)\citenamefont {Helsen}, \citenamefont {Ioannou}, \citenamefont {Kitzinger}, \citenamefont {Onorati}, \citenamefont {Werner}, \citenamefont {Eisert},\ and\ \citenamefont {Roth}}]{helsen2021estimating}%
  \BibitemOpen
  \bibfield  {author} {\bibinfo {author} {\bibfnamefont {J.}~\bibnamefont {Helsen}}, \bibinfo {author} {\bibfnamefont {M.}~\bibnamefont {Ioannou}}, \bibinfo {author} {\bibfnamefont {J.}~\bibnamefont {Kitzinger}}, \bibinfo {author} {\bibfnamefont {E.}~\bibnamefont {Onorati}}, \bibinfo {author} {\bibfnamefont {A.}~\bibnamefont {Werner}}, \bibinfo {author} {\bibfnamefont {J.}~\bibnamefont {Eisert}},\ and\ \bibinfo {author} {\bibfnamefont {I.}~\bibnamefont {Roth}},\ }\bibfield  {title} {\bibinfo {title} {Shadow estimation of gate-set properties from random sequences},\ }\href {https://doi.org/10.1038/s41467-023-39382-9} {\bibfield  {journal} {\bibinfo  {journal} {Nature Communications}\ }\textbf {\bibinfo {volume} {14}},\ \bibinfo {pages} {5039} (\bibinfo {year} {2023})}\BibitemShut {NoStop}%
\bibitem [{\citenamefont {White}\ \emph {et~al.}(2023)\citenamefont {White}, \citenamefont {Modi},\ and\ \citenamefont {Hill}}]{white2023filtering}%
  \BibitemOpen
  \bibfield  {author} {\bibinfo {author} {\bibfnamefont {G.~A.~L.}\ \bibnamefont {White}}, \bibinfo {author} {\bibfnamefont {K.}~\bibnamefont {Modi}},\ and\ \bibinfo {author} {\bibfnamefont {C.~D.}\ \bibnamefont {Hill}},\ }\bibfield  {title} {\bibinfo {title} {Filtering crosstalk from bath non-{M}arkovianity via spacetime classical shadows},\ }\href {https://doi.org/10.1103/PhysRevLett.130.160401} {\bibfield  {journal} {\bibinfo  {journal} {Phys. Rev. Lett.}\ }\textbf {\bibinfo {volume} {130}},\ \bibinfo {pages} {160401} (\bibinfo {year} {2023})}\BibitemShut {NoStop}%
\bibitem [{\citenamefont {Ippoliti}\ and\ \citenamefont {Khemani}(2024)}]{ippoliti2024learnability}%
  \BibitemOpen
  \bibfield  {author} {\bibinfo {author} {\bibfnamefont {M.}~\bibnamefont {Ippoliti}}\ and\ \bibinfo {author} {\bibfnamefont {V.}~\bibnamefont {Khemani}},\ }\bibfield  {title} {\bibinfo {title} {Learnability transitions in monitored quantum dynamics via eavesdropper's classical shadows},\ }\href {https://doi.org/10.1103/PRXQuantum.5.020304} {\bibfield  {journal} {\bibinfo  {journal} {PRX Quantum}\ }\textbf {\bibinfo {volume} {5}},\ \bibinfo {pages} {020304} (\bibinfo {year} {2024})}\BibitemShut {NoStop}%
\bibitem [{\citenamefont {Ruiz~Guzman}\ and\ \citenamefont {Lacroix}(2024)}]{ruiz2024restoring}%
  \BibitemOpen
  \bibfield  {author} {\bibinfo {author} {\bibfnamefont {E.~A.}\ \bibnamefont {Ruiz~Guzman}}\ and\ \bibinfo {author} {\bibfnamefont {D.}~\bibnamefont {Lacroix}},\ }\bibfield  {title} {\bibinfo {title} {Restoring symmetries in quantum computing using classical shadows},\ }\href {https://doi.org/10.1140/epja/s10050-024-01314-6} {\bibfield  {journal} {\bibinfo  {journal} {The European Physical Journal A}\ }\textbf {\bibinfo {volume} {60}},\ \bibinfo {pages} {1} (\bibinfo {year} {2024})}\BibitemShut {NoStop}%
\bibitem [{\citenamefont {Conrad}\ \emph {et~al.}(2024)\citenamefont {Conrad}, \citenamefont {Eisert},\ and\ \citenamefont {Flammia}}]{conrad2024chasing}%
  \BibitemOpen
  \bibfield  {author} {\bibinfo {author} {\bibfnamefont {J.}~\bibnamefont {Conrad}}, \bibinfo {author} {\bibfnamefont {J.}~\bibnamefont {Eisert}},\ and\ \bibinfo {author} {\bibfnamefont {S.~T.}\ \bibnamefont {Flammia}},\ }\bibfield  {title} {\bibinfo {title} {Chasing shadows with {G}ottesman-{K}itaev-{P}reskill codes},\ }\href {https://doi.org/10.48550/arXiv.2411.00235} {\bibfield  {journal} {\bibinfo  {journal} {arXiv preprint arXiv:2411.00235}\ } (\bibinfo {year} {2024})}\BibitemShut {NoStop}%
\bibitem [{\citenamefont {Chen}\ \emph {et~al.}(2021)\citenamefont {Chen}, \citenamefont {Yu}, \citenamefont {Zeng},\ and\ \citenamefont {Flammia}}]{chen2020robust}%
  \BibitemOpen
  \bibfield  {author} {\bibinfo {author} {\bibfnamefont {S.}~\bibnamefont {Chen}}, \bibinfo {author} {\bibfnamefont {W.}~\bibnamefont {Yu}}, \bibinfo {author} {\bibfnamefont {P.}~\bibnamefont {Zeng}},\ and\ \bibinfo {author} {\bibfnamefont {S.~T.}\ \bibnamefont {Flammia}},\ }\bibfield  {title} {\bibinfo {title} {{Robust Shadow Estimation}},\ }\href {https://doi.org/10.1103/PRXQuantum.2.030348} {\bibfield  {journal} {\bibinfo  {journal} {PRX Quantum}\ }\textbf {\bibinfo {volume} {2}},\ \bibinfo {pages} {030348} (\bibinfo {year} {2021})}\BibitemShut {NoStop}%
\bibitem [{\citenamefont {Koh}\ and\ \citenamefont {Grewal}(2022)}]{kohClassicalShadowsNoise2022}%
  \BibitemOpen
  \bibfield  {author} {\bibinfo {author} {\bibfnamefont {D.~E.}\ \bibnamefont {Koh}}\ and\ \bibinfo {author} {\bibfnamefont {S.}~\bibnamefont {Grewal}},\ }\bibfield  {title} {\bibinfo {title} {Classical {{Shadows With Noise}}},\ }\href {https://doi.org/10.22331/q-2022-08-16-776} {\bibfield  {journal} {\bibinfo  {journal} {Quantum}\ }\textbf {\bibinfo {volume} {6}},\ \bibinfo {pages} {776} (\bibinfo {year} {2022})}\BibitemShut {NoStop}%
\bibitem [{\citenamefont {Brieger}\ \emph {et~al.}(2023)\citenamefont {Brieger}, \citenamefont {Heinrich}, \citenamefont {Roth},\ and\ \citenamefont {Kliesch}}]{brieger2023stability}%
  \BibitemOpen
  \bibfield  {author} {\bibinfo {author} {\bibfnamefont {R.}~\bibnamefont {Brieger}}, \bibinfo {author} {\bibfnamefont {M.}~\bibnamefont {Heinrich}}, \bibinfo {author} {\bibfnamefont {I.}~\bibnamefont {Roth}},\ and\ \bibinfo {author} {\bibfnamefont {M.}~\bibnamefont {Kliesch}},\ }\bibfield  {title} {\bibinfo {title} {Stability of classical shadows under gate-dependent noise},\ }\href {https://doi.org/10.48550/arXiv.2310.19947} {\bibfield  {journal} {\bibinfo  {journal} {arXiv preprint arXiv:2310.19947}\ } (\bibinfo {year} {2023})}\BibitemShut {NoStop}%
\bibitem [{\citenamefont {Nguyen}(2023)}]{nguyen2023shadow}%
  \BibitemOpen
  \bibfield  {author} {\bibinfo {author} {\bibfnamefont {H.-C.}\ \bibnamefont {Nguyen}},\ }\bibfield  {title} {\bibinfo {title} {Shadow tomography with noisy readouts},\ }\href {https://doi.org/10.48550/arXiv.2310.17328} {\bibfield  {journal} {\bibinfo  {journal} {arXiv preprint arXiv:2310.17328}\ } (\bibinfo {year} {2023})}\BibitemShut {NoStop}%
\bibitem [{\citenamefont {Wu}\ and\ \citenamefont {Koh}(2024)}]{wuErrormitigatedFermionicClassical2024}%
  \BibitemOpen
  \bibfield  {author} {\bibinfo {author} {\bibfnamefont {B.}~\bibnamefont {Wu}}\ and\ \bibinfo {author} {\bibfnamefont {D.~E.}\ \bibnamefont {Koh}},\ }\bibfield  {title} {\bibinfo {title} {Error-mitigated fermionic classical shadows on noisy quantum devices},\ }\href {https://doi.org/10.1038/s41534-024-00836-7} {\bibfield  {journal} {\bibinfo  {journal} {npj Quantum Information}\ }\textbf {\bibinfo {volume} {10}},\ \bibinfo {pages} {39} (\bibinfo {year} {2024})}\BibitemShut {NoStop}%
\bibitem [{\citenamefont {Rozon}\ \emph {et~al.}(2024)\citenamefont {Rozon}, \citenamefont {Bao},\ and\ \citenamefont {Agarwal}}]{rozon2024optimal}%
  \BibitemOpen
  \bibfield  {author} {\bibinfo {author} {\bibfnamefont {P.-G.}\ \bibnamefont {Rozon}}, \bibinfo {author} {\bibfnamefont {N.}~\bibnamefont {Bao}},\ and\ \bibinfo {author} {\bibfnamefont {K.}~\bibnamefont {Agarwal}},\ }\bibfield  {title} {\bibinfo {title} {Optimal twirling depth for classical shadows in the presence of noise},\ }\href {https://doi.org/10.1103/PhysRevLett.133.130803} {\bibfield  {journal} {\bibinfo  {journal} {Physical Review Letters}\ }\textbf {\bibinfo {volume} {133}},\ \bibinfo {pages} {130803} (\bibinfo {year} {2024})}\BibitemShut {NoStop}%
\bibitem [{\citenamefont {Jnane}\ \emph {et~al.}(2024)\citenamefont {Jnane}, \citenamefont {Steinberg}, \citenamefont {Cai}, \citenamefont {Nguyen},\ and\ \citenamefont {Koczor}}]{jnane2024quantum}%
  \BibitemOpen
  \bibfield  {author} {\bibinfo {author} {\bibfnamefont {H.}~\bibnamefont {Jnane}}, \bibinfo {author} {\bibfnamefont {J.}~\bibnamefont {Steinberg}}, \bibinfo {author} {\bibfnamefont {Z.}~\bibnamefont {Cai}}, \bibinfo {author} {\bibfnamefont {H.~C.}\ \bibnamefont {Nguyen}},\ and\ \bibinfo {author} {\bibfnamefont {B.}~\bibnamefont {Koczor}},\ }\bibfield  {title} {\bibinfo {title} {Quantum error mitigated classical shadows},\ }\href {https://doi.org/10.1103/PRXQuantum.5.010324} {\bibfield  {journal} {\bibinfo  {journal} {PRX Quantum}\ }\textbf {\bibinfo {volume} {5}},\ \bibinfo {pages} {010324} (\bibinfo {year} {2024})}\BibitemShut {NoStop}%
\bibitem [{\citenamefont {Farias}\ \emph {et~al.}(2024)\citenamefont {Farias}, \citenamefont {Peddinti}, \citenamefont {Roth},\ and\ \citenamefont {Aolita}}]{farias2024robust}%
  \BibitemOpen
  \bibfield  {author} {\bibinfo {author} {\bibfnamefont {R.}~\bibnamefont {Farias}}, \bibinfo {author} {\bibfnamefont {R.~D.}\ \bibnamefont {Peddinti}}, \bibinfo {author} {\bibfnamefont {I.}~\bibnamefont {Roth}},\ and\ \bibinfo {author} {\bibfnamefont {L.}~\bibnamefont {Aolita}},\ }\bibfield  {title} {\bibinfo {title} {Robust shallow shadows},\ }\href {https://doi.org/10.48550/arXiv.2405.06022} {\bibfield  {journal} {\bibinfo  {journal} {arXiv preprint arXiv:2405.06022}\ } (\bibinfo {year} {2024})}\BibitemShut {NoStop}%
\bibitem [{\citenamefont {Onorati}\ \emph {et~al.}(2024)\citenamefont {Onorati}, \citenamefont {Kitzinger}, \citenamefont {Helsen}, \citenamefont {Ioannou}, \citenamefont {Werner}, \citenamefont {Roth},\ and\ \citenamefont {Eisert}}]{onorati2024noise}%
  \BibitemOpen
  \bibfield  {author} {\bibinfo {author} {\bibfnamefont {E.}~\bibnamefont {Onorati}}, \bibinfo {author} {\bibfnamefont {J.}~\bibnamefont {Kitzinger}}, \bibinfo {author} {\bibfnamefont {J.}~\bibnamefont {Helsen}}, \bibinfo {author} {\bibfnamefont {M.}~\bibnamefont {Ioannou}}, \bibinfo {author} {\bibfnamefont {A.}~\bibnamefont {Werner}}, \bibinfo {author} {\bibfnamefont {I.}~\bibnamefont {Roth}},\ and\ \bibinfo {author} {\bibfnamefont {J.}~\bibnamefont {Eisert}},\ }\bibfield  {title} {\bibinfo {title} {Noise-mitigated randomized measurements and self-calibrating shadow estimation},\ }\href {https://doi.org/10.48550/arXiv.2403.04751} {\bibfield  {journal} {\bibinfo  {journal} {arXiv preprint arXiv:2403.04751}\ } (\bibinfo {year} {2024})}\BibitemShut {NoStop}%
\bibitem [{\citenamefont {Devroye}\ \emph {et~al.}(2016)\citenamefont {Devroye}, \citenamefont {Lerasle}, \citenamefont {Lugosi},\ and\ \citenamefont {Oliveira}}]{devroyeSubGaussianMeanEstimators2015}%
  \BibitemOpen
  \bibfield  {author} {\bibinfo {author} {\bibfnamefont {L.}~\bibnamefont {Devroye}}, \bibinfo {author} {\bibfnamefont {M.}~\bibnamefont {Lerasle}}, \bibinfo {author} {\bibfnamefont {G.}~\bibnamefont {Lugosi}},\ and\ \bibinfo {author} {\bibfnamefont {R.~I.}\ \bibnamefont {Oliveira}},\ }\bibfield  {title} {\bibinfo {title} {Sub-{G}aussian mean estimators},\ }\href {http://www.jstor.org/stable/44245766} {\bibfield  {journal} {\bibinfo  {journal} {The Annals of Statistics}\ }\textbf {\bibinfo {volume} {44}},\ \bibinfo {pages} {2695} (\bibinfo {year} {2016})}\BibitemShut {NoStop}%
\bibitem [{\citenamefont {Lugosi}\ and\ \citenamefont {Mendelson}(2019)}]{lugosiMeanEstimationRegression2019}%
  \BibitemOpen
  \bibfield  {author} {\bibinfo {author} {\bibfnamefont {G.}~\bibnamefont {Lugosi}}\ and\ \bibinfo {author} {\bibfnamefont {S.}~\bibnamefont {Mendelson}},\ }\bibfield  {title} {\bibinfo {title} {Mean {{Estimation}} and {{Regression Under Heavy-Tailed Distributions}}: {{A Survey}}},\ }\href {https://doi.org/10.1007/s10208-019-09427-x} {\bibfield  {journal} {\bibinfo  {journal} {Foundations of Computational Mathematics}\ }\textbf {\bibinfo {volume} {19}},\ \bibinfo {pages} {1145} (\bibinfo {year} {2019})}\BibitemShut {NoStop}%
\bibitem [{\citenamefont {Struchalin}\ \emph {et~al.}(2021)\citenamefont {Struchalin}, \citenamefont {Zagorovskii}, \citenamefont {Kovlakov}, \citenamefont {Straupe},\ and\ \citenamefont {Kulik}}]{struchalinExperimentalEstimationQuantum2021}%
  \BibitemOpen
  \bibfield  {author} {\bibinfo {author} {\bibfnamefont {G.}~\bibnamefont {Struchalin}}, \bibinfo {author} {\bibfnamefont {{\relax Ya}.~A.}\ \bibnamefont {Zagorovskii}}, \bibinfo {author} {\bibfnamefont {E.}~\bibnamefont {Kovlakov}}, \bibinfo {author} {\bibfnamefont {S.}~\bibnamefont {Straupe}},\ and\ \bibinfo {author} {\bibfnamefont {S.}~\bibnamefont {Kulik}},\ }\bibfield  {title} {\bibinfo {title} {Experimental {{Estimation}} of {{Quantum State Properties}} from {{Classical Shadows}}},\ }\href {https://doi.org/10.1103/PRXQuantum.2.010307} {\bibfield  {journal} {\bibinfo  {journal} {PRX Quantum}\ }\textbf {\bibinfo {volume} {2}},\ \bibinfo {pages} {010307} (\bibinfo {year} {2021})}\BibitemShut {NoStop}%
\bibitem [{\citenamefont {Zhang}\ \emph {et~al.}(2021)\citenamefont {Zhang}, \citenamefont {Sun}, \citenamefont {Fang}, \citenamefont {Zhang}, \citenamefont {Yuan},\ and\ \citenamefont {Lu}}]{zhangExperimentalQuantumState2021}%
  \BibitemOpen
  \bibfield  {author} {\bibinfo {author} {\bibfnamefont {T.}~\bibnamefont {Zhang}}, \bibinfo {author} {\bibfnamefont {J.}~\bibnamefont {Sun}}, \bibinfo {author} {\bibfnamefont {X.-X.}\ \bibnamefont {Fang}}, \bibinfo {author} {\bibfnamefont {X.-M.}\ \bibnamefont {Zhang}}, \bibinfo {author} {\bibfnamefont {X.}~\bibnamefont {Yuan}},\ and\ \bibinfo {author} {\bibfnamefont {H.}~\bibnamefont {Lu}},\ }\bibfield  {title} {\bibinfo {title} {Experimental {{Quantum State Measurement}} with {{Classical Shadows}}},\ }\href {https://doi.org/10.1103/PhysRevLett.127.200501} {\bibfield  {journal} {\bibinfo  {journal} {Physical Review Letters}\ }\textbf {\bibinfo {volume} {127}},\ \bibinfo {pages} {200501} (\bibinfo {year} {2021})}\BibitemShut {NoStop}%
\bibitem [{\citenamefont {Levy}\ \emph {et~al.}(2024)\citenamefont {Levy}, \citenamefont {Luo},\ and\ \citenamefont {Clark}}]{levyClassicalShadowsQuantum2024}%
  \BibitemOpen
  \bibfield  {author} {\bibinfo {author} {\bibfnamefont {R.}~\bibnamefont {Levy}}, \bibinfo {author} {\bibfnamefont {D.}~\bibnamefont {Luo}},\ and\ \bibinfo {author} {\bibfnamefont {B.~K.}\ \bibnamefont {Clark}},\ }\bibfield  {title} {\bibinfo {title} {Classical shadows for quantum process tomography on near-term quantum computers},\ }\href {https://doi.org/10.1103/PhysRevResearch.6.013029} {\bibfield  {journal} {\bibinfo  {journal} {Physical Review Research}\ }\textbf {\bibinfo {volume} {6}},\ \bibinfo {pages} {013029} (\bibinfo {year} {2024})}\BibitemShut {NoStop}%
\bibitem [{\citenamefont {Dutt}\ \emph {et~al.}(2023)\citenamefont {Dutt}, \citenamefont {Kirby}, \citenamefont {Raymond}, \citenamefont {Hadfield}, \citenamefont {Sheldon}, \citenamefont {Chuang},\ and\ \citenamefont {Mezzacapo}}]{duttPracticalBenchmarkingRandomized2023}%
  \BibitemOpen
  \bibfield  {author} {\bibinfo {author} {\bibfnamefont {A.}~\bibnamefont {Dutt}}, \bibinfo {author} {\bibfnamefont {W.}~\bibnamefont {Kirby}}, \bibinfo {author} {\bibfnamefont {R.}~\bibnamefont {Raymond}}, \bibinfo {author} {\bibfnamefont {C.}~\bibnamefont {Hadfield}}, \bibinfo {author} {\bibfnamefont {S.}~\bibnamefont {Sheldon}}, \bibinfo {author} {\bibfnamefont {I.~L.}\ \bibnamefont {Chuang}},\ and\ \bibinfo {author} {\bibfnamefont {A.}~\bibnamefont {Mezzacapo}},\ }\bibfield  {title} {\bibinfo {title} {{Practical Benchmarking of Randomized Measurement Methods for Quantum Chemistry Hamiltonians}},\ }\href {https://doi.org/10.48550/arXiv.2312.07497} {\bibfield  {journal} {\bibinfo  {journal} {arXiv preprint arXiv:2312.07497}\ } (\bibinfo {year} {2023})}\BibitemShut {NoStop}%
\bibitem [{\citenamefont {Minsker}(2023{\natexlab{a}})}]{minskerEfficientMedianMeans2023}%
  \BibitemOpen
  \bibfield  {author} {\bibinfo {author} {\bibfnamefont {S.}~\bibnamefont {Minsker}},\ }\bibfield  {title} {\bibinfo {title} {Efficient median of means estimator},\ }in\ \href {https://proceedings.mlr.press/v195/minsker23a.html} {\emph {\bibinfo {booktitle} {Proceedings of Thirty Sixth Conference on Learning Theory}}},\ \bibinfo {series} {Proceedings of Machine Learning Research}, Vol.\ \bibinfo {volume} {195},\ \bibinfo {editor} {edited by\ \bibinfo {editor} {\bibfnamefont {G.}~\bibnamefont {Neu}}\ and\ \bibinfo {editor} {\bibfnamefont {L.}~\bibnamefont {Rosasco}}}\ (\bibinfo  {publisher} {PMLR},\ \bibinfo {year} {2023})\ pp.\ \bibinfo {pages} {5925--5933}\BibitemShut {NoStop}%
\bibitem [{\citenamefont {Minsker}(2023{\natexlab{b}})}]{minskerUstatisticsGrowingOrder2023}%
  \BibitemOpen
  \bibfield  {author} {\bibinfo {author} {\bibfnamefont {S.}~\bibnamefont {Minsker}},\ }\bibfield  {title} {\bibinfo {title} {U-statistics of growing order and sub-{{Gaussian}} mean estimators with sharp constants},\ }\href {https://doi.org/10.4171/msl/43} {\bibfield  {journal} {\bibinfo  {journal} {Mathematical Statistics and Learning}\ }\textbf {\bibinfo {volume} {7}},\ \bibinfo {pages} {1} (\bibinfo {year} {2023}{\natexlab{b}})}\BibitemShut {NoStop}%
\bibitem [{\citenamefont {Lee}(2019)}]{leeUStatisticsTheoryPractice2019}%
  \BibitemOpen
  \bibfield  {author} {\bibinfo {author} {\bibfnamefont {A.~J.}\ \bibnamefont {Lee}},\ }\href {https://doi.org/10.1201/9780203734520} {\emph {\bibinfo {title} {U-{{Statistics}}: {{Theory}} and {{Practice}}}}}\ (\bibinfo  {publisher} {Routledge},\ \bibinfo {address} {New York},\ \bibinfo {year} {2019})\BibitemShut {NoStop}%
\bibitem [{\citenamefont {Greenberger}\ \emph {et~al.}(1989)\citenamefont {Greenberger}, \citenamefont {Horne},\ and\ \citenamefont {Zeilinger}}]{greenberger1989going}%
  \BibitemOpen
  \bibfield  {author} {\bibinfo {author} {\bibfnamefont {D.~M.}\ \bibnamefont {Greenberger}}, \bibinfo {author} {\bibfnamefont {M.~A.}\ \bibnamefont {Horne}},\ and\ \bibinfo {author} {\bibfnamefont {A.}~\bibnamefont {Zeilinger}},\ }\bibinfo {title} {Going beyond bell's theorem},\ in\ \href {https://doi.org/10.1007/978-94-017-0849-4_10} {\emph {\bibinfo {booktitle} {Bell's Theorem, Quantum Theory and Conceptions of the Universe}}},\ \bibinfo {editor} {edited by\ \bibinfo {editor} {\bibfnamefont {M.}~\bibnamefont {Kafatos}}}\ (\bibinfo  {publisher} {Springer Netherlands},\ \bibinfo {address} {Dordrecht},\ \bibinfo {year} {1989})\ pp.\ \bibinfo {pages} {69--72}\BibitemShut {NoStop}%
\bibitem [{\citenamefont {Minsker}(2019)}]{minskerDistributedStatisticalEstimation2019}%
  \BibitemOpen
  \bibfield  {author} {\bibinfo {author} {\bibfnamefont {S.}~\bibnamefont {Minsker}},\ }\bibfield  {title} {\bibinfo {title} {Distributed statistical estimation and rates of convergence in normal approximation},\ }\href {https://doi.org/10.1214/19-EJS1647} {\bibfield  {journal} {\bibinfo  {journal} {Electronic Journal of Statistics}\ }\textbf {\bibinfo {volume} {13}},\ \bibinfo {pages} {5213} (\bibinfo {year} {2019})}\BibitemShut {NoStop}%
\bibitem [{\citenamefont {Drakakis}(2009)}]{drakakisReviewAvailableConstruction2009}%
  \BibitemOpen
  \bibfield  {author} {\bibinfo {author} {\bibfnamefont {K.}~\bibnamefont {Drakakis}},\ }\bibfield  {title} {\bibinfo {title} {A review of the available construction methods for {G}olomb rulers.},\ }\href {https://doi.org/10.3934/amc.2009.3.235} {\bibfield  {journal} {\bibinfo  {journal} {Advances in Mathematics of Communications}\ }\textbf {\bibinfo {volume} {3}},\ \bibinfo {pages} {235} (\bibinfo {year} {2009})}\BibitemShut {NoStop}%
\bibitem [{\citenamefont {Ferris}\ and\ \citenamefont {Vidal}(2012)}]{ferrisPerfectSamplingUnitary2012}%
  \BibitemOpen
  \bibfield  {author} {\bibinfo {author} {\bibfnamefont {A.~J.}\ \bibnamefont {Ferris}}\ and\ \bibinfo {author} {\bibfnamefont {G.}~\bibnamefont {Vidal}},\ }\bibfield  {title} {\bibinfo {title} {Perfect sampling with unitary tensor networks},\ }\href {https://doi.org/10.1103/PhysRevB.85.165146} {\bibfield  {journal} {\bibinfo  {journal} {Physical Review B}\ }\textbf {\bibinfo {volume} {85}},\ \bibinfo {pages} {165146} (\bibinfo {year} {2012})}\BibitemShut {NoStop}%
\bibitem [{\citenamefont {Huang}\ \emph {et~al.}(2021{\natexlab{b}})\citenamefont {Huang}, \citenamefont {Kueng},\ and\ \citenamefont {Preskill}}]{huangEfficientEstimationPauli2021}%
  \BibitemOpen
  \bibfield  {author} {\bibinfo {author} {\bibfnamefont {H.-Y.}\ \bibnamefont {Huang}}, \bibinfo {author} {\bibfnamefont {R.}~\bibnamefont {Kueng}},\ and\ \bibinfo {author} {\bibfnamefont {J.}~\bibnamefont {Preskill}},\ }\bibfield  {title} {\bibinfo {title} {Efficient {{Estimation}} of {{Pauli Observables}} by {{Derandomization}}},\ }\href {https://doi.org/10.1103/PhysRevLett.127.030503} {\bibfield  {journal} {\bibinfo  {journal} {Physical Review Letters}\ }\textbf {\bibinfo {volume} {127}},\ \bibinfo {pages} {030503} (\bibinfo {year} {2021}{\natexlab{b}})}\BibitemShut {NoStop}%
\end{thebibliography}%

\newpage
\appendix
\onecolumngrid
\section{MoM Bound}
\label{sec:mom_bound}
We present a proof for the bound
\begin{align}
    \label{eq:mom-bound}
    P[|\hat \mu_N - \expec(X)| \geq \e] \leq 2e^{-k/2},
\end{align}
where \(\e = \sqrt{{4e^2\sigma^2 k}/{N}}\) and \(N\)  is the total number of data points for an MoM estimator \(\hat \mu_N\), adapted from Devroye et al.~\cite{devroyeSubGaussianMeanEstimators2015}.

Given random variables \(X_1, \ldots, X_N\), we split them into \(k\) disjoint sets of size \(b = \lfloor N / k \rfloor\), and find their mean. If \(|\hat \mu_N - \expec(X)| \geq \e\), this means more than half of the means \(\bar X_i^{(b)} = b^{-1} \sum_{j \in J_i} X_j,\ |J_i| = b \) deviates from \(\expec(X)\) by more than \(\e\). Let \(Y^{(b)}_i = \bar X_i^{(b)} - \expec(X)\):
\begin{align}
    \sum_{i=1}^{k} I\left\{|Y^{(b)}_i| \geq \e\right\} &\geq \frac{k}{2}.
\end{align}
As \(Y^{(b)}_i\) are i.i.d., let \(\var\left[X_i\right] = \sigma^2 < \infty\) and \(\var\left[Y^{(b)}_i\right] = k\sigma^2/N\) so that by Chebyshev's inequality,
\begin{equation}
     P(|Y^{(b)}_i| \geq \e) \leq \frac{k\sigma^2}{N\e^2}.
\end{equation}
\(|Y^{(l)}| \geq \e\) is a Bernoulli event with probability \(\leq{k\sigma^2}/{N\e^2}\). This implies
\begin{equation}
    P\left(\sum_{i=1}^{k} I\left\{|Y^{(b)}_i| \geq \e\right\} = r\right) \leq { k\choose r} \left(\frac{k\sigma^2}{N\e^2}\right)^r \left(1-\frac{k\sigma^2}{N\e^2}\right)^{k-r}.
\end{equation}

Putting everything together,
\begin{align}
    P(|\hat \mu_N - \expec(X)| \geq \e) &\leq P\left( \sum_{i=1}^{k} I\left\{|Y^{(b)}_i| \geq \e\right\} \geq \frac{k}{2} \right)\\
    &\leq \sum_{r=\lceil k/2\rceil}^k { k\choose r} \left(\frac{k\sigma^2}{N\e^2}\right)^r \left(1-\frac{k\sigma^2}{N\e^2}\right)^{k-r}\\
    &\leq \left(\frac{k\sigma^2}{N\e^2}\right)^{\lceil k/2\rceil} \sum_{r=\lceil k/2\rceil}^k { k\choose r}\\
     &\leq \left(\frac{k\sigma^2}{N\e^2}\right)^{\lceil k/2\rceil} \sum_{r=0}^k { k\choose r}\\
     &= \left(\frac{k\sigma^2}{N\e^2}\right)^{\lceil k/2\rceil} 2^k.
\end{align}
Choosing \(N = 4 e^2 \sigma^2 k / \e^2\) ensures that
\begin{align}
    P(|\hat \mu_N - \expec(X)| \geq \e)&\leq e^{-k}\\
    &\leq 2e^{-k/2},
\end{align}
thereby obtaining the bound given in \autoref{eq:mom-bound}.

\section{Bounds for Gaussian Distribution}
\label{sec:gaussian_bounds}

In the simulations, the mean performed better despite having worse bounds than the MoM estimators when using Chebyshev's inequality (\autoref{eq:mean_bound}). However, for large shadow size, the distribution of the estimator \(\hat o_i(N, 1)\) tends towards a Gaussian distribution, with \(P(|\hat \mu - \mu | \geq \e) \leq \exp\ab(-\frac{N \e^2}{2 \sigma^2}) \equiv \frac{\delta}{M}\), giving an explanation for the better-than-expected performance of the mean. Here, we include comparisons between this tighter bound under this normality assumption, and the bound in \autoref{thm:new_bound}.

\begin{figure}
    \centering
    \includegraphics[width=0.5\linewidth]{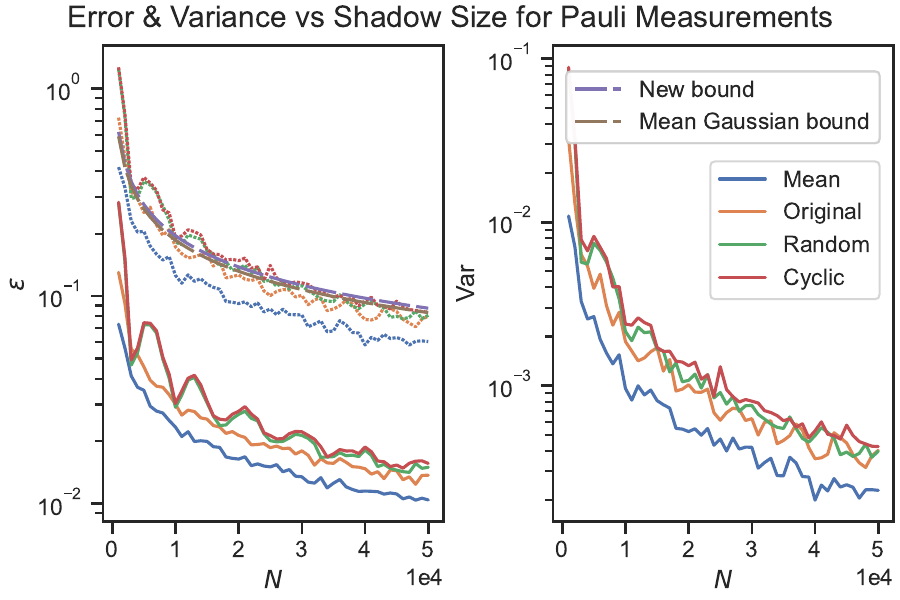}
    \caption{Error and variance vs shadow size for the Ising model using Pauli measurements when estimating the correlator, as in \autoref{fig:ising_pauli_error}. We compare the bounds under the assumption that the distribution of the mean of the shadow is Gaussian, and the bound by Minsker in \autoref{thm:new_bound}.}
    \label{fig:ising_pauli_error_chernoff}
\end{figure}

\begin{figure}
    \centering
    \includegraphics[width=0.5\linewidth]{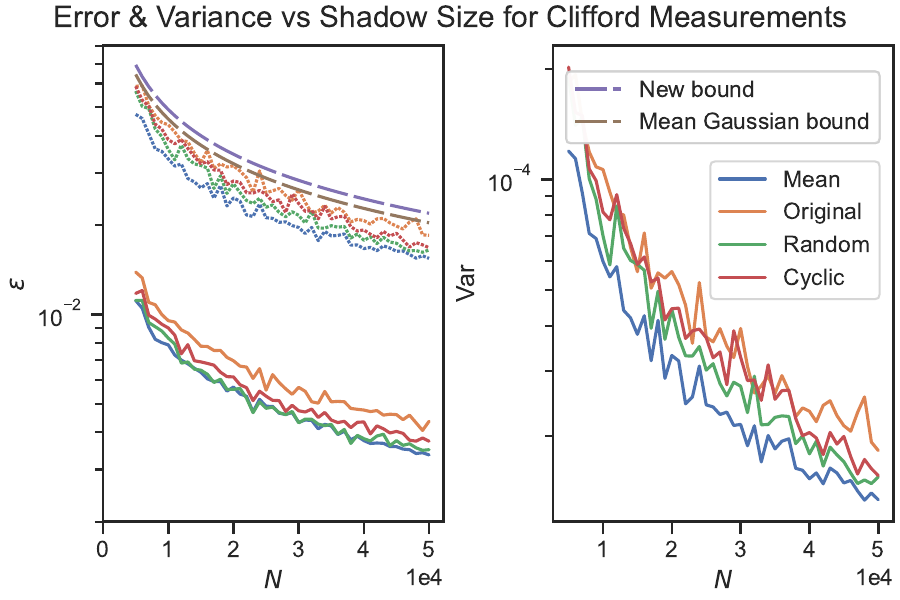}
    \caption{Error and variance vs shadow size for the 2-qubit GHZ state using Clifford measurements when estimating fidelity, as in \autoref{fig:ghz_error}, with comparison of bounds under the normality assumption.}
    \label{fig:ghz_error_chernoff}
\end{figure}

\begin{figure}
    \centering
    \includegraphics[width=0.5\linewidth]{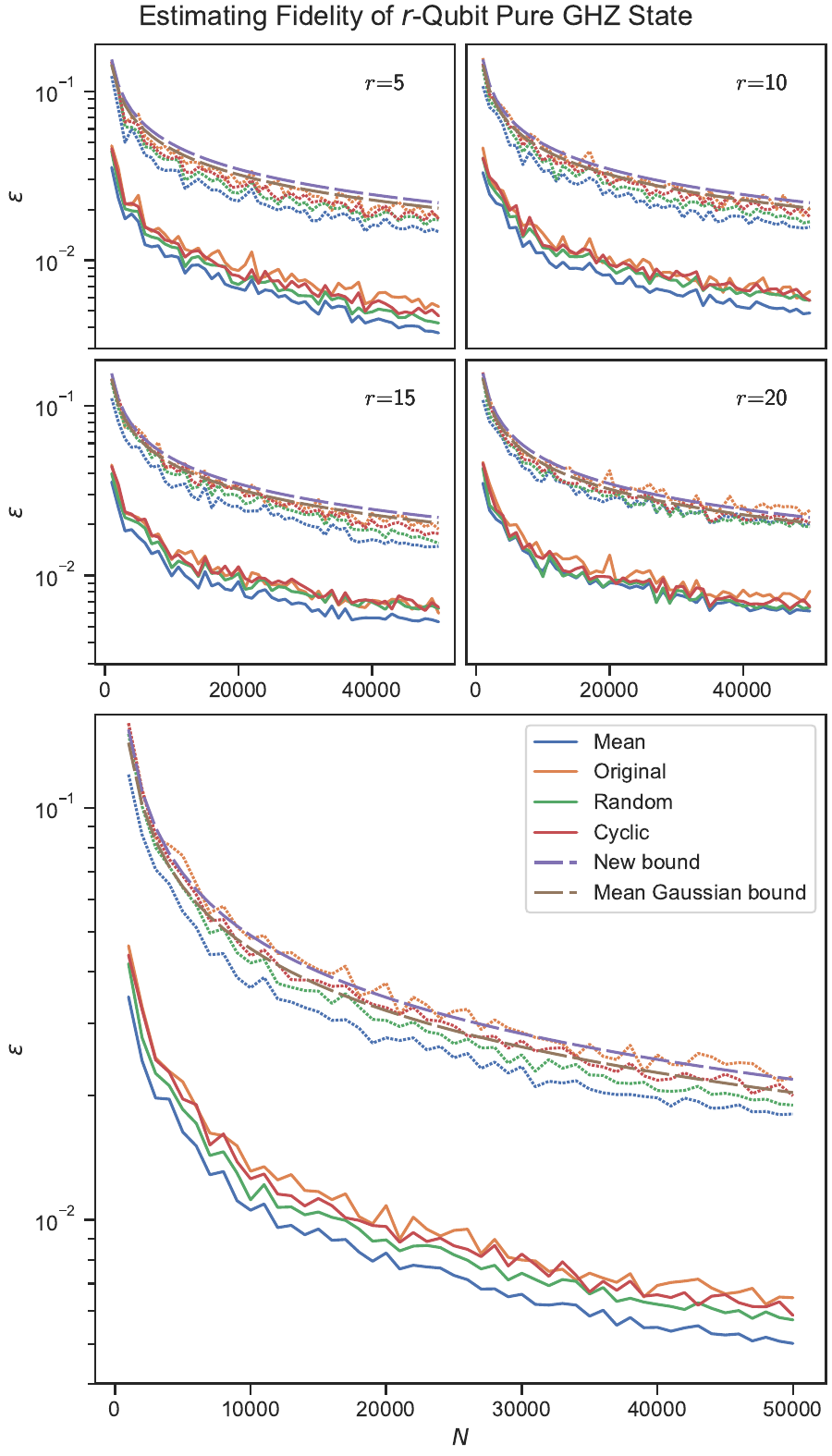}
    \caption{Error and variance vs shadow size for the 5, 10, 15, 20-qubit GHZ states using Clifford measurements when estimating fidelity, as in \autoref{fig:ghz_fid_var_qubits}, with comparison of bounds under the normality assumption.}
    \label{fig:ghz_fid_var_qubits_chernoff}
\end{figure}

\begin{figure}
    \centering
    \includegraphics[width=0.5\linewidth]{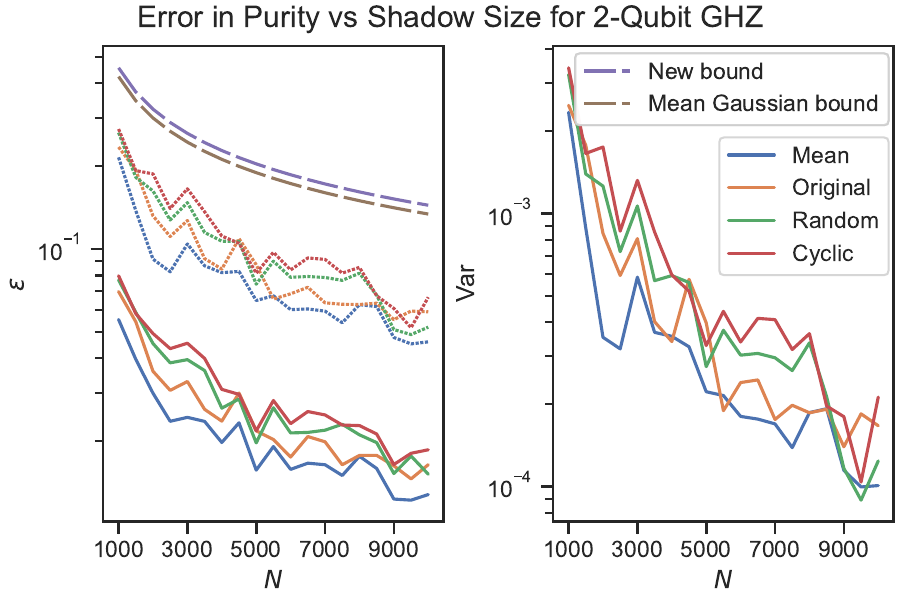}
    \caption{Error and variance vs shadow size for the 2-qubit GHZ state using Clifford measurements when estimating the purity, as in \autoref{fig:ghz_purity_error}, with comparison of bounds under the normality assumption.}
    \label{fig:ghz_purity_error_chernoff}
\end{figure}

This `Mean Gaussian bound' is tighter than \autoref{thm:new_bound} and obeyed by the mean for the Ising model (\autoref{fig:ising_pauli_error_chernoff}), 2-qubit GHZ (\autoref{fig:ghz_error_chernoff}), \(5, 10, 15, 20\)-qubit GHZ (\autoref{fig:ghz_fid_var_qubits_chernoff}), and GHZ purity (\autoref{fig:ghz_purity_error_chernoff}). In \autoref{fig:ghz_fid_var_qubits_chernoff}, the tighter bound is obeyed by the mean and \ref{func:momrand}, but not \ref{func:mom} and \ref{func:momcyc}. These results demonstrate the validity of the normality assumption for classical shadows.

\section{Performance of Estimators for Pauli Observables}
\label{sec:hit_count}

The performance of the estimators may be explained by the different values of \(k\) used for each, as suggested by analysis by Huang et al.\ \cite{huangEfficientEstimationPauli2021}. The mean is equivalent to \ref{func:mom} with \(k=1\). For \ref{func:mom}, \(k = 43\). For \ref{func:momrand} and \ref{func:momcyc}, \(k = [65, 219]\) when \(N = [1000,50000]\). \(k\) changes the number of data points \(\trace(O_i \hat \rho_j)\) that the final estimate is averaged over, corresponding to \(\lfloor N/k \rfloor\).

\begin{figure}
    \centering
    \includegraphics[width=0.4\linewidth]{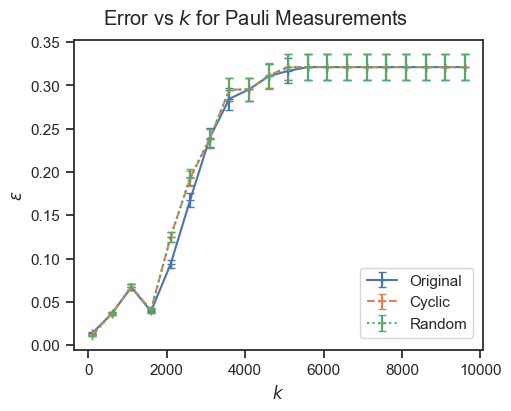}
    \caption{Error vs group size \(k\) for predicting the 2-point correlator for 50 qubits, using different estimators, averaged over 10 independent runs.}
    \label{fig:ising_k_error}
\end{figure}

\autoref{fig:ising_k_error} shows that, rather than the type of estimator used, the error depends most heavily on the number of groups \(k\). The fewer data points per group and the higher \(k\) used, the higher the error and variance of the estimator. 

In the aforementioned paper, Huang et al. showed that the failure probability of the estimator for Pauli observables and measurements is exponentially suppressed by the `hit count' of the estimator. For a system of \(n_q\) qubits, we perform \(N\) Pauli measurements \(p_i \in \{X, Y, Z\}^{n_q} \), to estimate \(M\) observables \(O_j\), with each measurement corresponding to a snapshot \(\hat \rho_i\). For Pauli observables, \autoref{eq:pauli_shadow_form} implies that
\(\trace(\hat \rho_i O_j)\) is non-zero if the measurement `hits' the observable, \(O_j \triangleright p_i\). That is, by substituting \(\{X, Y, Z\}\) in \(p_i\) with \(I\), we can obtain \(O_j\) if \(O_j \triangleright p_i\). Huang et al.\ showed that the performance of the estimator improves exponentially with the number of hits. By splitting the data set into groups, the hit count in each group is reduced, exponentially increasing the failure probability of each \(\bar X_i\) for \ref{func:mom} and \(\bar Z_J\) for \ref{func:momrand} and \ref{func:momcyc}.

\end{document}